%% file: main.tex
\documentclass[sigconf]{acmart}

\usepackage{booktabs}
\usepackage{graphicx}
\usepackage{listings}
\usepackage{xcolor}
\usepackage{amsmath}
\usepackage{enumitem}
\usepackage{subcaption}
\usepackage{multirow}
\usepackage{balance}
\usepackage{pifont}
\usepackage{tcolorbox}
\usepackage{multicol}
\usepackage{float}
\usepackage{wrapfig}

\begin{document}

\title{Quality-Assured Fuzz Harness Generation via\\the Four Principles Framework}

\author{Ze Sheng}
\affiliation{\institution{Texas A\&M University}\country{USA}}
\email{zesheng@tamu.edu}

\author{Dmitrijs Trizna}
\affiliation{\institution{Aisle}\country{USA}}
\email{dimi@aisle.com}

\author{Luigino Camastra}
\affiliation{\institution{Aisle}\country{USA}}
\email{luigino.camastra@aisle.com}

\author{Zhicheng Chen}
\affiliation{\institution{Texas A\&M University}\country{USA}}
\email{chenzc2001@tamu.edu}

\author{Qingxiao Xu}
\affiliation{\institution{Texas A\&M University}\country{USA}}
\email{qingxiao@tamu.edu}

\author{Jeff Huang}
\affiliation{\institution{Texas A\&M University}\country{USA}}
\email{jeffhuang@tamu.edu}

\begin{abstract}
Fuzz testing is the dominant technique for finding memory-safety vulnerabilities in C/C++ software, yet its effectiveness hinges on the quality of \emph{fuzz harnesses}---the programs that bridge fuzzers and library APIs.
A growing body of tools now automate harness generation, but none systematically ensures the \emph{correctness} of produced harnesses: logic errors, API misuse, and lifecycle violations go undetected at the source level, producing false-positive crash reports at rates as high as 94\%.
As LLM-driven generation scales harness creation, uncontrolled quality turns scale into a liability.

We present \emph{QuartetFuzz}, an autonomous harness-generation system that systematically improves correctness throughout the generation process.
At its core is the \emph{Four Principles} framework---Logic Correctness~(P1), API Protocol Compliance~(P2), Security Boundary Respect~(P3), and Entry Point Adequacy~(P4)---the first source-level definition of harness correctness with mathematical specifications and implementable checks.
We operationalize these principles in an autonomous LLM agent that produces harnesses satisfying P1--P4 through a generate--check--fix loop before any fuzzing begins.

Deployed on 23~open-source projects spanning C/C++, Java, and JavaScript, the system submits \textbf{42~bug reports}, of which \textbf{29~are fixed or confirmed upstream} (including 3 CVEs) and only 2~are rejected (\textbf{4.8\% FP rate}).
During generation, the built-in P1/P2 checks automatically intercepted 58 harness-induced crashes that would otherwise have been false positives.
Applied as a quality auditor to 586~existing production harnesses across 70~projects, the system identifies \textbf{53~violations} (\textbf{45~confirmed}, \textbf{35~fixed}).
We release a dataset of \textbf{100~labeled harnesses} for reproducible evaluation. Code and dataset are available at \url{https://github.com/OwenSanzas/QuartetFuzz}.
\end{abstract}



\maketitle

\input{sections/introduction}

\input{sections/background}

\input{sections/system_design}
\input{sections/implementation}
\input{sections/evaluation}

\input{sections/discussion}
\input{sections/related_work}
\input{sections/conclusion}

\input{main.bbl}

\input{sections/ccs_meta}

\input{sections/appendix}

\end{document}

%% file: sections/introduction.tex
\section{Introduction}
\label{sec:intro}

Fuzz testing~\cite{miller1990fuzz} is one of the most effective techniques for discovering memory-safety vulnerabilities in C/C++ software, but for libraries its success depends on the quality of the \emph{fuzz harness}: the code that translates raw fuzzer bytes into library API calls.
A harness does more than make the target compile.
It decides which functionality is exercised, how deeply the fuzzer can reach, and whether observed crashes reflect real library bugs or mistakes in the harness.
A wrong harness undermines the campaign's effectiveness; the crashes it produces become hard to interpret.

This problem has become more severe as automated harness generation has improved.
Traditional systems such as FuzzGen~\cite{fuzzgen}, FUDGE~\cite{fudge}, Hopper~\cite{hopper}, and AFGen~\cite{afgen}, and more recent LLM-based systems such as OSS-Fuzz-Gen~\cite{ossfuzzgen}, PromptFuzz~\cite{promptfuzz}, CKGFuzzer~\cite{ckgfuzzer}, PromeFuzz~\cite{promefuzz}, and HarnessAgent~\cite{harnessagent} all aim to reduce the effort of writing harnesses and to improve fuzzing reach.
Yet they still evaluate quality largely through proxy signals such as build success, crash counts, and code coverage;
the only prior pre-generation checker, Scheduzz~\cite{scheduzz}, validates types, not API protocol.
Those signals are necessary, but they are not sufficient: a harness can compile, run, and even reach substantial code coverage while still violating API contracts, masking deep code paths, or producing false-positive crashes---at rates as high as 94\% in prior work~\cite{nexzzer}.
As LLMs make harness generation easy to scale, uncontrolled harness quality turns that scale into a liability.

\begin{figure}[t]
\centering
\includegraphics[width=\columnwidth]{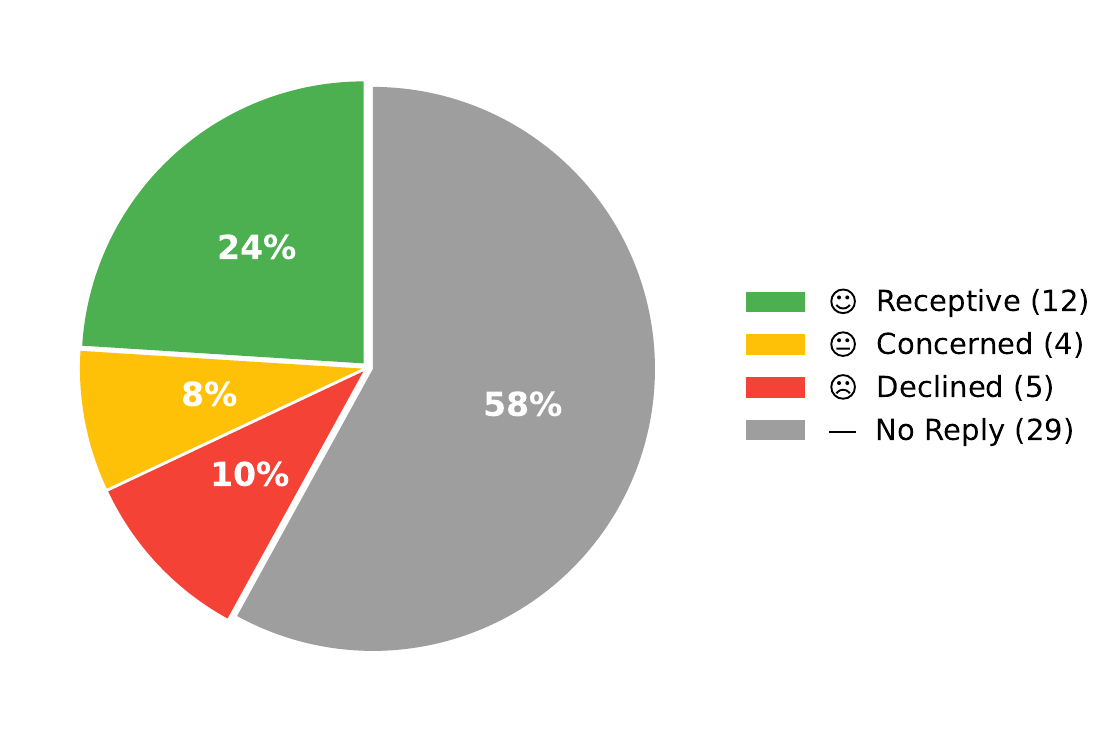}
\caption{Developer responses when contacted about AI-assisted harness generation across 50 OSS-Fuzz projects. 58\% never responded; among respondents, reactions ranged from receptive (24\%) to outright refusal (10\%), with concerns centered on false-positive overload from prior automated tools.}
\label{fig:developer_survey}
\end{figure}

We saw this concern directly in practice.
When we contacted maintainers of 50 OSS-Fuzz projects about collaborating on AI-generated harnesses and bug reports, the dominant concern was not generation speed or raw coverage but trust: maintainers had prior experience with automated fuzzing workflows that produced large numbers of false-positive reports whose sanitizer traces looked legitimate, but whose root cause was a defect in the harness rather than the library (Figure~\ref{fig:developer_survey}).
In other words, the cost of poor harness quality is not only wasted CPU time;
it is shifted triage burden for maintainers.
This motivates a different question from most prior work: not merely whether a harness runs, but whether it is \emph{correct}.

To investigate, we manually audited 586 production OSS-Fuzz harnesses across 70 C/C++ projects (\S\ref{sec:eval:rq1}); informal reviews of harnesses from LLM4FDG~\cite{llm4fdg}, PromptFuzz~\cite{promptfuzz}, and PromeFuzz~\cite{promefuzz} showed the same patterns.
From fuzzers exhibiting abnormal coverage, false-positive crashes, or runtime errors, we identified four correctness principles a harness must satisfy:
\emph{P1, Logic Correctness}---the harness itself must be free of bugs such as stale state, resource leaks, or incorrect data handling;
\emph{P2, API Protocol Compliance}---the harness must call the target library in a valid order with valid object lifecycles and parameter constraints;
\emph{P3, Security Boundary Respect}---the harness should exercise public interfaces rather than bypassing validation through internal-only code;
and
\emph{P4, Entry Point Adequacy}---the harness should target meaningful, attack-surface-relevant entry points rather than incidental helper logic.
We call these the \textbf{Four Principles} of fuzz harness quality.

Guided by these principles, we build \emph{QuartetFuzz}, an LLM-based harness generation system that checks P1--P4 during generation, before any fuzzing campaign begins.
The system operates in four stages: (1) an \emph{entry-selection agent} uses the call graph to enumerate candidate fuzzing targets, ensures each can reach its core implementation, prefers public APIs, and ranks them by a static \emph{danger score};
(2) an \emph{API-research agent} collects protocol information (call order, object lifecycle, parameter constraints) for the selected targets;
(3) a \emph{harness-generator agent} drafts a fuzzer source in full and runs a bounded build loop;
and (4) an \emph{adversarial-validation gate} forces the harness through reach- and run-check probes, routing any crash through P1/P2/P3/P4 triage before submission.
Prior generators---including LLM-based ones---target individual functions in isolation, reducing the LLM to a wrapper writer despite its demonstrated ability to comprehend code semantics across files~\cite{lprotector}.
Our system instead organizes fuzzing around \emph{Logic Groups}---functional units (e.g., ``parse a PNG image,'' ``complete a TLS handshake'') that capture stateful interactions across related APIs.

Our results show that harness quality is not a secondary nicety but a practical bottleneck.
Applied as a quality auditor to production harnesses, QuartetFuzz identifies 53 quality violations;
45 have been confirmed by maintainers and 35 have been fixed or merged upstream.
Repairing those defects exposed 2 latent library bugs the harnesses had been masking, including a stack-buffer-overread in OpenSSL's DES implementation that had been latent for over 25~years.
On a dataset of 100 gold-standard harnesses curated from the audit (39 projects, all verified P1--P4 clean), the generator matches human-written harnesses on coverage (TOST $\pm2$pp equivalence, $p{<}10^{-10}$ on both line and branch) while outperforming OSS-Fuzz-Gen by 6.9--8.3pp and PromeFuzz by 4.1--5.2pp across line and branch coverage.
Deployed on 23 open-source projects spanning C/C++, Java, and JavaScript, QuartetFuzz submitted 42 bug reports: 29 are fixed or confirmed upstream (including 3 CVEs) and only 2 were rejected as false positives (4.8\% FP rate).
Across audit and generation, the same P1/P2 checks blocked \textbf{58 harness-induced crashes} (14 in audited production harnesses, 44 in generated ones) before they could become false-positive bug reports.
The 35 merged repairs and four follow-up adoptions support the paper's central claim: harness quality should be checked \emph{before} fuzzing begins, not inferred afterward from crash logs.
In summary, this paper makes the following contributions.

\noindent\textbf{The Four Principles framework with Adversarial Probing.}
We define harness correctness as source-level conditions (P1--P4), operationalised by \emph{Adversarial Probing} (P1/P2 runtime sub-checks) and static call-graph reachability (P3/P4 boundary sub-checks).
To our knowledge, no prior LLM-based generator turns the agent on its own output before deployment.

\noindent\textbf{End-to-end pipeline producing deliverable harnesses.}
We present a four-stage LLM-agent pipeline (Logic Group selection, API protocol research, static-driven build, adversarial validation) that produces quality-assured harnesses ready for direct upstream merge.
Applied as a post-hoc auditor on 586 production OSS-Fuzz harnesses, the same module identified 53 quality violations of which 45 are confirmed and 35 merged upstream---each is a deliverable harness running in or queued for production OSS-Fuzz.

\noindent\textbf{A curated evaluation dataset.}
From the 586 audited production harnesses, we select 100 P1--P4-clean harnesses across 39 projects as a gold-standard dataset for evaluating harness generation against real-world developer-written fuzzers.
The dataset and evaluation pipeline are released for reproducible comparison.

\noindent\textbf{Real-world bug discovery.}
Deployed on 23 open-source projects spanning C/C++, Java, and JavaScript, the pipeline yielded 42 bug reports (29 fixed or confirmed, 3 CVEs, 4.8\% FP);
audit repairs additionally exposed 2 latent library bugs, including an OpenSSL DES stack-buffer-overread latent for over 25 years, and 4 maintainer teams integrated our harnesses or derived artifacts.

%% file: sections/background.tex
\section{Background}
\label{sec:background}

\subsection{Fuzz Testing and Fuzz Harnesses}
\label{sec:bg:fuzzing}

Coverage-guided fuzz testing~\cite{klees2018evaluating} repeatedly feeds mutated inputs to a program, using code coverage feedback to explore new execution paths.
Fuzzers such as libFuzzer~\cite{libfuzzer} and AFL++~\cite{aflpp}, typically paired with sanitizers such as AddressSanitizer~\cite{asan2012} and LeakSanitizer~\cite{lsan}, are standard tools for discovering memory-safety vulnerabilities in C/C++ software.
For standalone programs, the fuzzer can target the main entry point directly.
For \emph{library} code, which exposes APIs but lacks a standalone entry point, a \emph{fuzz harness} (also called a fuzz driver or fuzz target) must be written.
A libFuzzer harness implements the following interface:

\begin{lstlisting}[basicstyle=\footnotesize\ttfamily]
int LLVMFuzzerTestOneInput(const uint8_t *data,
                           size_t size) {
    // parse data, call library APIs, clean up
    return 0;
}
\end{lstlisting}

\noindent
The fuzzer engine calls this function repeatedly with different byte sequences.
The harness is responsible for (1)~parsing raw bytes into typed arguments, (2)~invoking library APIs in a meaningful sequence, (3)~handling errors without crashing, and (4)~releasing all allocated resources before returning.

Google's OSS-Fuzz platform~\cite{ossfuzz} continuously fuzzes over 1,000 open-source projects with handwritten harnesses.
We focus on C/C++ library harnesses in OSS-Fuzz throughout.

\subsection{Automated Harness Generation}
\label{sec:bg:generation}

Fuzz harness generation can be formulated as a program synthesis problem.
The inputs are a codebase $\mathcal{L}$ with API set $\mathcal{A} = \{a_1, \ldots, a_n\}$, and a target specification $\tau$ at a chosen granularity---a single API function, a co-called API group, or a higher-level functionality description (e.g., ``PNG decoding'').
The task is to synthesize a program $H$ that implements the fuzzer interface, maps raw bytes to typed arguments, and composes a call sequence over a subset $S \subseteq \mathcal{A}$ that exercises $\tau$.

A valid harness must satisfy $H \models \mathcal{C}(\mathcal{L}, \tau)$, where $\mathcal{C}$ is the set of conditions $H$ must pass to be accepted as a correct harness for $\tau$ in $\mathcal{L}$.
Both the granularity at which $\tau$ is specified and the contents of $\mathcal{C}$ vary by methodology.

Existing tools fix the granularity of $\tau$---single API in OSS-Fuzz-Gen~\cite{ossfuzzgen} and HarnessAgent~\cite{harnessagent}, API clusters in PromptFuzz~\cite{promptfuzz}, CKGFuzzer~\cite{ckgfuzzer}, PromeFuzz~\cite{promefuzz}, and Scheduzz~\cite{scheduzz}, internal target functions in pre-LLM tools~\cite{fuzzgen,fudge,winnie,apicraft,utopia}---and run a generate--compile--fix--validate loop with $\mathcal{C} = \{\mathcal{C}_\text{call}, \mathcal{C}_\text{fuzz}, \mathcal{C}_\text{cov}\}$, where $\mathcal{C}_\text{call}$ requires $H$ to invoke the target function~\cite{harnessagent}, $\mathcal{C}_\text{fuzz}$ requires short-duration fuzzing (30\,s in OSS-Fuzz-Gen~\cite{ossfuzzgen}, 60\,s in LLM4FDG~\cite{llm4fdg}) not to surface crashes traceable to harness bugs, and $\mathcal{C}_\text{cov}$ requires $H$'s execution to increase coverage on the target.

%% file: sections/system_design.tex
\section{System Design}
\label{sec:design}

QuartetFuzz primarily targets C/C++ open-source projects, producing LibFuzzer harnesses;
Logic Group ranking consumes a precomputed call graph of the project.
The pipeline separates LLM reasoning from this call-graph input cleanly enough that the same flow applies to languages where the call graph is unavailable:
\S\ref{sec:eval:rq5} extends to Java and JavaScript with LLM-only Logic Group ranking.
The four principles in \S\ref{sec:design:principles} characterize harness quality;
the four-stage pipeline (\S\ref{sec:design:overview}) checks all four before fuzzing.

\subsection{The Four Principles}
\label{sec:design:principles}

We extend the validation set $\mathcal{C}$ from \S\ref{sec:bg:generation} with four source-level conditions, P1--P4, checked statically and dynamically before deployment fuzzing.
Each yields a pass/fail verdict;
a harness is accepted only when all four pass.

\noindent\textbf{Setup and notation.}
P1 and P2 are runtime properties of $H$.
P3 and P4 are call-graph properties (\S\ref{sec:design:lg}).
$H$ targets a specification $\tau$, given as a Logic Group with entry set $E_\tau \subseteq \mathcal{A}$ and core set $C_\tau \subseteq \mathcal{A}$.
Let $\mathcal{A}$ be the library API set and $\mathcal{B}$ the space of fuzz blobs.
A fuzz blob sequence $\vec{b} = (b_1, \ldots, b_N) \in \mathcal{B}^N$ drives $H(\vec{b}) := H(b_1);
\ldots;
H(b_N)$ in one shared process (the libFuzzer model).
For each $b_i$, $H$ produces a trace $T_H(b_i) = \langle (a_1, \textit{args}_1), \ldots, (a_k, \textit{args}_k) \rangle$, the prefix of calls (possibly empty) issued before $H$ crashes or returns.
$\textsc{direct}(T)$ replays $T$ in a fresh process with the exact arguments the library originally observed and no harness-level parsing or state.
$\textsc{direct}$ is undefined when $\textit{args}_j$ depends on harness-internal state;
\S\ref{sec:impl} approximates it via per-iteration sub-process isolation.

P1 uses two fault oracles.
$O_\text{single}$ detects faults that show up within a single trace;
$O_\text{seq}$ detects faults that show up only across iterations.
We instantiate $O_\text{single}$ as AddressSanitizer~\cite{asan2012} and $O_\text{seq}$ as AddressSanitizer + LeakSanitizer~\cite{lsan}.
$\textsc{crash}_O(\cdot)$ is true when oracle $O$ reports a fault on the given execution.
In this work we assume execution is deterministic, so flaky crashes are out of scope.

Both $\textsc{direct}$ and $O_\text{single}$ serve as semantic references and are not literally executed;
$O_\text{seq}$ is realized by libFuzzer with ASan and LSan in the build--fix loop.

\noindent\textbf{Principle 1: Logic Correctness (P1).}
P1 requires
\begin{equation}
\begin{aligned}
\forall \vec{b}:\; & \big(\forall i \in \{1, \ldots, N\}:\, \neg \textsc{crash}_{O_\text{single}}(\textsc{direct}(T_H(b_i)))\big) \\
& \;\Longrightarrow\; \neg \textsc{crash}_{O_\text{seq}}(H(\vec{b})).
\end{aligned}
\end{equation}
The implication separates harness faults from per-call API behavior;
this requires asymmetric oracles, since an end-of-process oracle (LSan) in $O_\text{single}$ would vacuously violate the antecedent on any leaking trace.
Faults that propagate through API arguments escape the mathematical property and are caught by the syntactic checks P1.1--P1.8 (\S\ref{sec:design:stage4}).

\noindent\textbf{Principle 2: API Protocol Compliance (P2).}
Each library function $f$ has a usage protocol $P_f$ with language $L(P_f)$ over its protocol-related operations $\mathrm{deps}(f) \subseteq \mathcal{A}$ (initialisers, parameter allocators, paired cleanups, return consumers).
$\mathrm{calls}(T)$ and $T|_S$ denote the standard call set and order-preserving projection.
P2 requires
\begin{equation}
\forall \vec{b}, \forall i, \forall f \in \mathrm{calls}(T_H(b_i)):\;
  T_H(b_i)|_{\mathrm{deps}(f)} \in L(P_f).
\end{equation}
$L(P_f)$ is partial.
\S\ref{sec:design:stage2} approximates it from headers, comments, and source.
\S\ref{sec:design:stage4} realizes membership as eight checks P2.1--P2.8 (init, parameter construction, lifecycle, return handling, cleanup, API existence, co-call, prerequisite state).
P2 is a property of the trace, independent of crashes;
when a trace violates both P1 and P2, we attribute to P2.

\noindent\textbf{Principle 3: Security Boundary Respect (P3).}
Let $\textsc{Pub} \subseteq \mathcal{A}$ be the public API surface and $\mathcal{H}_{\textsc{Pub}}$ the set of P2-respecting public-only harnesses.
$\textsc{crash}_O(X, A)$ holds when oracle $O$ fires at library code site $A$ (function, line).
P3 requires every $H$ crash to be reproducible by some public-only harness at the same site:
\begin{equation}
\begin{aligned}
\forall \vec{b}, A:\; & \textsc{crash}_O(H(\vec{b}), A) \\
& \;\Longrightarrow\; \exists H' \in \mathcal{H}_{\textsc{Pub}},\, \vec{b}':\; \textsc{crash}_O(H'(\vec{b}'), A).
\end{aligned}
\end{equation}
A P3 violation is a crash reachable only via internal entries that bypass library defenses;
no public-only harness can reproduce it, so it falls outside the threat model.
P3 is not statically decidable;
static reachability gives a necessary condition, and \S\ref{sec:design:stage1} details the fallback when no public entry reaches $C_\tau$.

\noindent\textbf{Principle 4: Entry Point Adequacy (P4).}
Let $\rightsquigarrow$ denote reachability in the call graph and $\textsc{unsafeReach}(e)$ a predicate that holds when $e$'s reach closure contains memory-safety-relevant operations.
P4 requires every fuzz-consuming entry to reach the target core along a path that exposes unsafe operations:
\begin{equation}
\forall e \in E_\tau:\;
  e \rightsquigarrow C_\tau \;\land\; \textsc{unsafeReach}(e).
\end{equation}
The first conjunct rules out entries that cannot reach the target;
the second rules out entries whose reach closure contains no memory-safety-relevant operations (e.g., wrappers into logging or accessors).
We instantiate $\textsc{unsafeReach}(\cdot)$ as $\textit{danger}(\cdot) > 0$ using the depth-discounted unsafe-operation score from \S\ref{sec:design:stage1} (Eq.~\ref{eq:danger});
other scorings (e.g., static taint, sanitizer allow-lists) realize the same property.

\medskip
\noindent The four principles are ordered by scope.
P1 concerns the harness, P2 the harness--library interface, P3 the public versus internal boundary, and P4 the link to real-world usage.
The generator applies them in reverse (P4 first, P1 last) because checking code correctness on the wrong entry is wasted.
P1 and P2 (runtime) are operationalised by Adversarial Probing (\S\ref{sec:design:probing});
P3 by static call-graph reachability (\S\ref{sec:design:stage1});
P4 by static reachability and a Stage~4 reach check (\S\ref{sec:design:stage4}).

\begin{figure}[H]
\begin{tcolorbox}[colback=gray!8, colframe=gray!60,
  title=Adversarial Probing: Two Probe Operations]
\small
Given an input blob, AP runs the harness in one of two ways:\\[3pt]
\textbf{Probe 1 --- Reach check}\\
\quad blob $\to$ run binary under GDB with breakpoint at target API\\
\quad $\to$ \textsc{hit}/\textsc{miss} + functions actually called\\
\quad \textbf{Targets:} P2 sub-checks needing observed call/lifecycle\\[3pt]
\textbf{Probe 2 --- Run check}\\
\quad blob $\to$ run binary under libFuzzer with ASan/LSan\\
\quad $\to$ pass/fail + sanitizer trace on fault\\
\quad \textbf{Targets:} P1 logic-correctness sub-checks\\[3pt]
\textit{Adversarial framing: the agent reads its own harness, picks the P1.x / P2.x sub-checks (e.g., P1.4 input flow, P1.6 size guard, P2.1 init order, P2.4 return value, P2.8 prerequisite state) the code looks most likely to violate, and writes a blob aimed at triggering each. Stage 4 admits the harness only if every selected probe passes.}
\end{tcolorbox}
\caption{The two probing operations of Adversarial Probing.}
\label{fig:ap_modes}
\end{figure}

\subsection{Adversarial Probing}
\label{sec:design:probing}

The mathematical definitions of P1 and P2 resist direct execution:
$\textsc{direct}$ is undefined when $\textit{args}_j$ depends on harness state, and $L(P_f)$ is only partially knowable from sources.
We bridge this gap with \emph{Adversarial Probing} (AP): two probe operations on input blobs (Figure~\ref{fig:ap_modes})---a \emph{reach check} (does the blob drive execution into the target API?) and a \emph{run check} (does the harness crash under ASan/LSan?).
The agent supplies adversarial blobs by writing code that generates a blob designed to trigger a suspected violation of a P1.x / P2.x sub-check;
libFuzzer also feeds inputs during the build--fix loop.
Four P1/P2 sub-checks resist probing (a single input cannot trigger them) and are decided statically (Table~\ref{tab:probe_attribution}).
AP fires only in Stage 4 (\S\ref{sec:design:stage4});
Stages 1--3 use source-level analysis only, mirroring how a human author first studies an API before testing.
P1.x and P2.x are therefore not LLM-as-checklist box-ticking but specific targets that probes attempt to violate.
The agent reads its harness, selects from Table~\ref{tab:probe_attribution} the sub-checks the code looks most likely to violate, and gives each one attack attempt;
the Stage 4 submission gate requires every selected probe to pass (no harness-bug crash) before the harness is eligible.
To our knowledge, no prior LLM-based generator inspects its harness's call sequence or constructs adversarial inputs against itself during generation.

\begin{figure*}[t]
\centering
\includegraphics[width=0.85\textwidth]{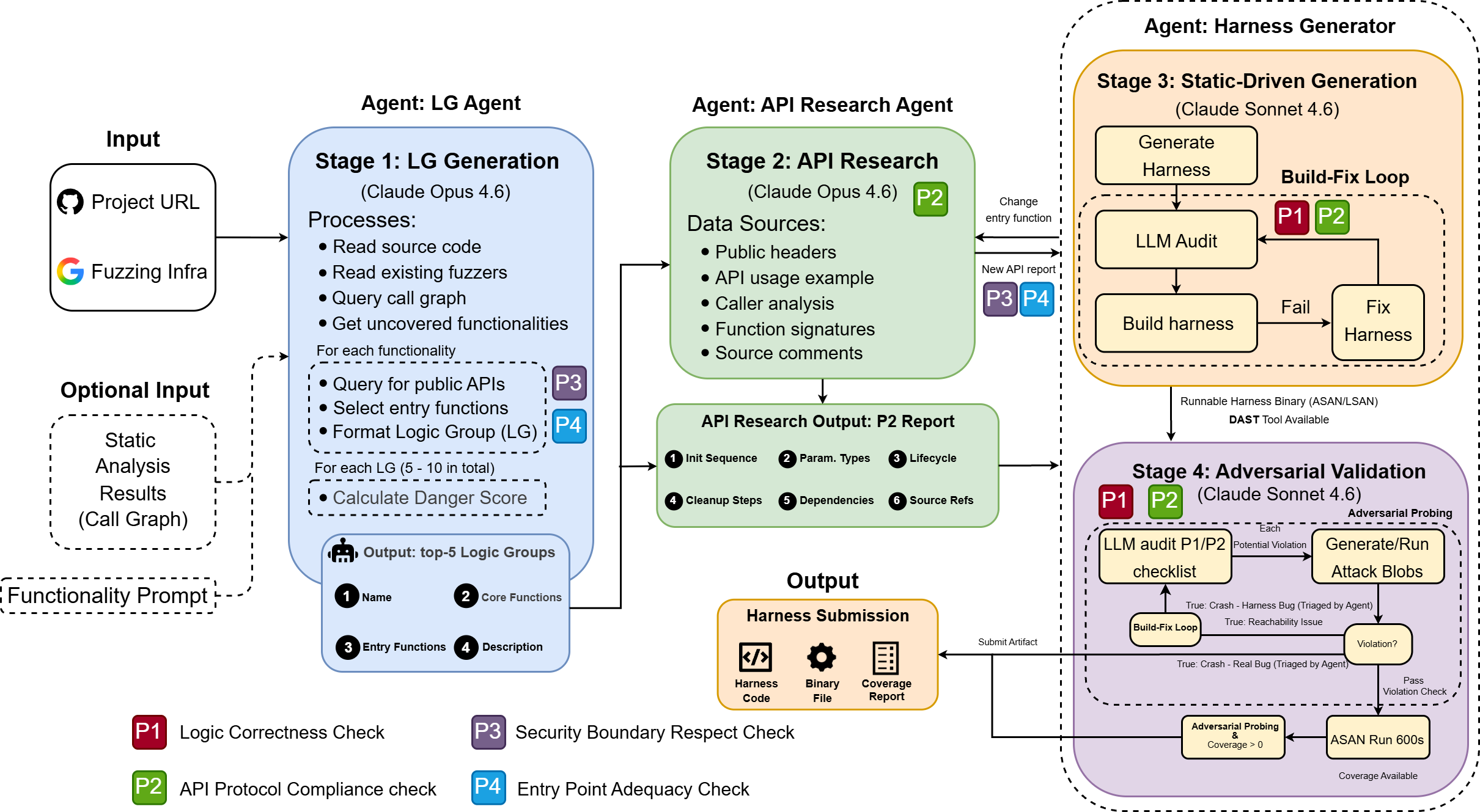}
\caption{System overview.  Stage~1: the Logic Group agent explores
the project, identifies candidate functionalities, and ranks entry
points by danger score (P3/P4 checked).  For each selected LG,
Stage~2 researches API protocol from headers and callers (P2).
Stage~3 produces a compilable binary via an LLM-drafted source
plus a bounded build loop.  Stage~4 puts that binary through
libFuzzer and agent-selected adversarial probes; any crash is
triaged as harness bug (fix), boundary issue (re-select entry),
or real vulnerability (submit upstream).}
\label{fig:overview}
\end{figure*}

\subsection{Logic Groups}
\label{sec:design:lg}

We organize fuzzing around \emph{features} rather than single functions.
A Logic Group $\text{LG} = (\text{name}, E, C, \text{desc})$ captures a feature.
$E$ is the entry set of library APIs receiving fuzzer-controlled bytes.
$C$ is a curated set of core functions transitively called from $E$ that implement the feature.
\emph{name} and \emph{desc} label the feature and describe its security relevance.
We write $\text{LG}_\tau$ with $E_\tau$, $C_\tau$ when emphasising a target $\tau$.
Figure~\ref{fig:lg_example} shows an example.

\begin{figure}[H]
\begin{tcolorbox}[colback=gray!8, colframe=gray!60, title=Logic Group: Transliteration rule processing (ICU)]
\small
\textbf{Entry functions:}\\
\quad \texttt{Transliterator::createFromRules} --- create transliterator from rule string\\[3pt]
\textbf{Core functions:}\\
\quad \texttt{TransliterationRule} --- parse and compile rules\\
\quad \texttt{TransliterationRuleParser} --- tokenize rule syntax\\
\quad \texttt{UnicodeSet::applyPattern} --- resolve character classes\\[3pt]
\textbf{Description:} Converts text between scripts using user-supplied transliteration rules; parses untrusted rule strings with complex syntax.\\[3pt]
\textit{Result: UAF vulnerability found (confirmed and fixed upstream), ranked \#1 by danger score among 5 candidates.}
\end{tcolorbox}
\caption{An example Logic Group generated by QuartetFuzz for ICU.}
\label{fig:lg_example}
\end{figure}

\subsection{System Overview}
\label{sec:design:overview}

Figure~\ref{fig:overview} shows the pipeline.
Stage~1 identifies 5--10 candidate Logic Groups, ranks them by danger, and selects the top~5.
Stage~2 collects API-protocol information for each selected LG.
Stage~3 (\emph{static-driven build}) uses only source reading and call-graph queries to produce a compilable binary.
Stage~4 (\emph{adversarial validation}) forces the harness through reach- and run-check probes plus crash triage before submission.
Stages~3 and~4 share a single Harness Generator agent;
the agent invokes its build, fuzz, and AP-probe tools without a strict step ordering---we do not enforce a fixed tool sequence between Stage~3 and Stage~4 (full tool list in Table~\ref{tab:agents}, Appendix~\ref{sec:appendix:agents}).

\subsection{Stage 1: Logic Group Discovery}
\label{sec:design:stage1}

The Logic Group agent reads the project source and generates 5--10 candidate Logic Groups, then selects the top~5.
P3 and P4 are the primary checks in this stage.
The pipeline runs in five steps.

\noindent\emph{Step 1: Existing coverage.}
If the project already has fuzzers, the agent reads them and reconstructs the LGs they cover, $\mathcal{G}_{\text{exist}}$, so subsequent candidates target uncovered ground.

\noindent\emph{Step 2: Candidate identification.}
The agent identifies 5--10 candidate features different from $\mathcal{G}_{\text{exist}}$.
For each candidate it sketches a tentative $C^{0}$ of 3--5 core functions (Figure~\ref{fig:lg_example}).

\noindent\emph{Step 3: Entry selection (P3/P4).}
For each candidate, the agent refines $C^{0}$ into $C$ by using SAST queries and source reading, dropping members that are not internal or central and adding adjacent helpers.
The agent then performs a reverse caller search (after AFGen~\cite{afgen}) to find public APIs reaching $C$:
\[
  E_{\text{pub}} = \{ f \in \mathcal{A}_{\text{pub}} \mid C \cap \text{callees}^{*}(f) \neq \emptyset \}.
\]
Equivalently, the agent walks callers backward from each $c \in C$ and keeps the public ancestors.
$\mathcal{A}_{\text{pub}}$ is built coarsely from path markers (\texttt{include/}, \texttt{public/}, \texttt{api/}) and refined by LLM judgment for ambiguous symbols (no clear marker---the agent reads the declaring header to decide).
If $E_{\text{pub}}\neq\emptyset$, P3 holds and the agent picks the entry with the shortest path to $C$ (P4 preference).
If $E_{\text{pub}}=\emptyset$, no public API reaches $C$;
the agent browses the call graph and picks an internal entry that reaches $C$ while preserving as many boundaries as possible (best-effort P3).

\noindent\emph{Step 4: Assembly and dedup.}
The agent assembles the LG and semantically deduplicates it against $\mathcal{G}_{\text{exist}}$ and previously-emitted LGs from prompt context.
When two new LGs share an entry, the LLM keeps both if they exercise distinct cores.

\noindent\emph{Step 5: Ranking.}
We rank the remaining LGs by a static danger score with $1/d$ depth discount, after IntelliGen~\cite{intelligen}:
\begin{equation}
\label{eq:danger}
\text{danger}(f) \;=\!\!
  \sum_{g \,\in\, \text{reachable}(f,\,D)}
  \frac{\text{unsafe}(g)}{d(f,\,g)}
\end{equation}
where $\text{reachable}(f, D)$ is the set of functions reachable
from~$f$ within $D$~hops on the static call graph ($D{=}20$;
$D{=}20$ sits on a stable plateau in our sensitivity analysis,
Appendix~\ref{sec:appendix:danger_sensitivity}),
$d(f,g)$ is the shortest-path distance from $f$ to~$g$, and
$\text{unsafe}(g)$ counts pointer dereferences and memory
operations (\texttt{memcpy}, \texttt{malloc}, \texttt{free},
\texttt{strcpy}, \texttt{sprintf}, etc.)\ in~$g$'s implementation.
By design, the $1/d$ factor concentrates the ranking on the directly-reachable attack surface---close unsafe operations that empty-corpus fuzzing can exercise without depending on a seed corpus.
When a Logic Group has multiple entry functions, its score is
$\text{danger}(\text{LG}) = \max_{f \in E}\,\text{danger}(f)$.
Stage~1 outputs the top-5 ranked Logic Groups for processing in
Stages~2 through~4.

\noindent\textbf{Why not rank by score alone?}
We avoid pure danger-score ranking for three reasons. (1) It wastes the LLM's code comprehension.
The model identifies parsers, protocol handlers, and state machines that no static metric captures.
(2) Greedy ranking biases toward large complex functions and skips simpler but security-critical code.
(3) Scoring all functions per project is slow.

\begin{figure*}[t]
\begin{tcolorbox}[colback=gray!8, colframe=gray!60, title=P2 Protocol Report: \texttt{Transliterator::createFromRules} (ICU)]
\scriptsize
\begin{multicols*}{2}
\textbf{P2.1 Init Sequence}\\
\textit{Claim:} No global init required. Call the static factory directly.
\texttt{UErrorCode} must be set to \texttt{U\_ZERO\_ERROR} before call.\\
\textit{Evidence:}\\
{\scriptsize\texttt{> i18n/unicode/translit.h:1109: @stable ICU 2.0}}\\
{\scriptsize\texttt{> test/intltest/transapi.cpp:671: direct call}}\\[3pt]

\textbf{P2.2 Parameter Construction}\\
\textit{Claim:}
\texttt{rules} (\texttt{UnicodeString}, semicolon-separated) carries
fuzz input; \texttt{dir} is \texttt{UTRANS\_FORWARD} or
\texttt{REVERSE}; \texttt{status} must be \texttt{U\_ZERO\_ERROR}.\\
\textit{Evidence:}\\
{\scriptsize\texttt{> i18n/unicode/translit.h:1103--1108: param docs}}\\[3pt]

\textbf{P2.3 Object Lifecycle}\\
\textit{Claim:}
1.~Create: returns \texttt{Transliterator*}; caller owns.
2.~Use: \texttt{t->transliterate(text)}.
3.~Destroy: \texttt{delete t} (virtual destructor).\\
\textit{Evidence:}\\
{\scriptsize\texttt{> i18n/translit.cpp:1062--1063: return nullptr on failure}}\\
{\scriptsize\texttt{> i18n/rbt\_rule.cpp:185--191: dtor frees 5 ptrs}}\\[3pt]

\textbf{P2.4 Return Value Handling}\\
\textit{Claim:} Success: valid pointer, must \texttt{delete}. Failure: \texttt{nullptr}.\\
\textit{Evidence:}\\
{\scriptsize\texttt{> i18n/translit.cpp:1062--1063: return nullptr on U\_FAILURE}}\\[3pt]

\columnbreak

\textbf{P2.5 Cleanup Sequence}\\
\textit{Claim:} \texttt{delete t} on success; no cleanup on failure.\\
\textit{Evidence:}\\
{\scriptsize\texttt{> test/intltest/transapi.cpp:84: delete t after use}}\\[3pt]

\textbf{P2.6 API Existence}\\
\textit{Claim:} All APIs verified to exist in ICU4C.\\
\textit{Evidence:}\\
{\scriptsize\texttt{> translit.h:1109: createFromRules (public)}}\\
{\scriptsize\texttt{> translit.h:652: transliterate (public)}}\\
{\scriptsize\texttt{> translit.h:611: \~{}Transliterator (virtual)}}\\[3pt]

\textbf{P2.7 Co-call Constraints}\\
\textit{Claim:} \texttt{createFromRules}/\texttt{delete} are paired;
check \texttt{U\_FAILURE} before using pointer.\\
\textit{Evidence:}\\
{\scriptsize\texttt{> i18n/translit.cpp:1062: early return before alloc}}\\[3pt]

\textbf{P2.8 Prerequisite State}\\
\textit{Claim:} None. No global init, locale, or I/O required.\\
\textit{Evidence:}\\
{\scriptsize\texttt{> i18n/translit.cpp:1051--1065: in-memory parser}}\\[6pt]

\textit{Note: the UAF vulnerability was in the destructor path
(P2.3) --- \texttt{rbt\_rule.cpp:185} deletes uninitialized
pointers when the constructor returns early.  This report enabled
the generator to write a harness that triggered the bug.}
\end{multicols*}
\end{tcolorbox}
\caption{Condensed P2 protocol report for ICU's
\texttt{createFromRules} API, generated by the API Research agent.
Each entry carries a project-specific \emph{claim} grounded in source-cited \emph{evidence} (file:line).
Sub-check definitions (P2.1--P2.8) are tabulated in Table~\ref{tab:probe_attribution}.}
\label{fig:p2_example}
\end{figure*}

\subsection{Stage 2: API Research}
\label{sec:design:stage2}

P2 violations (wrong call order, missing init, incorrect cleanup) are the dominant failure mode of both human and LLM harnesses (\S\ref{sec:eval:rq1}).
Most prior systems give the LLM only partial protocol context:
API correlation metrics over documentation~\cite{promefuzz}, function signatures from project introspection~\cite{ossfuzzgen}, or post-hoc crash-derived constraints~\cite{afgen}.
We instead run a dedicated API Research agent that pre-collects protocol information before the generator writes code.
With full read access to the project source, the agent works like a human library expert---performing semantic analysis over the codebase rather than relying on structural heuristics or signature introspection---and decides what to read on its own.

For each entry, the agent
(1) reads the public header for signature and preconditions,
(2) walks the call graph to find 2--3 production callers and reads their source,
(3) inspects internal dependencies through call-graph callees, and
(4) searches tests and examples.

Its output is a structured P2 protocol report covering eight sub-checks (P2.1--P2.8 defined in Table~\ref{tab:probe_attribution}); each entry carries a \emph{claim} (the agent's project-specific finding) and \emph{evidence} (file path and line).
Grounding every claim in a verifiable source location keeps the agent from inventing protocol rules.
Figure~\ref{fig:p2_example} shows a concrete instance for ICU's \texttt{createFromRules}.

\subsection{Stage 3: Static-Driven Build}
\label{sec:design:stage3}

Stages~3 and~4 run inside one Harness Generator agent: Stage~3 produces a binary that compiles, Stage~4 then attacks that binary with AP probes.

The agent first reads the project's build configuration (build script, Dockerfile, existing fuzzers, license) to settle on the right source style (\texttt{.c}, \texttt{.cc}, \texttt{.cpp}) and to match the conventions any existing fuzzers in the project follow.
It then reads the library source, consults the Stage~1 Logic Group and the Stage~2 P2 report, and drafts the fuzzer source.
While drafting, it reviews every P1/P2 sub-check via LLM semantic reasoning over the draft.
No binary exists yet, so all 16 sub-checks are evaluated as source-level inspections rather than by their runtime oracles in Table~\ref{tab:probe_attribution}.
It then calls build\_harness; on failure it reads the compiler tail, edits the source, and retries.
No AP probes and no fuzz run in Stage~3;
the output is a binary that compiles, at which point Stage~4 begins.

\subsection{Stage 4: Adversarial Validation}
\label{sec:design:stage4}

Stage~4 is an Adversarial Probing (AP)-driven loop that decides whether the binary is ready to submit.
The agent attacks its own harness to verify P1/P2 sub-checks.
For brevity we reuse the term \emph{build-fix loop} from Stage~3, but Stage~4's build-fix loop additionally has access to AP probes and get\_coverage (after the 600\,s ASan run).
The steps below detail the loop body.

\emph{Step 1: Adversarial probing.}
The agent reads its own harness together with the relevant library source, and selects from Table~\ref{tab:probe_attribution} any P1.x / P2.x sub-checks the code is most likely to violate (no fixed count), writing one adversarial blob per selected check.
AP\_run\_check / AP\_reach\_check execute the blobs one by one against the binary, aggregate the per-blob results, and return the aggregated output to the agent, which reviews all outcomes together.
Every outcome falls into one of four classes:
\begin{enumerate}[leftmargin=*]
\item \textbf{Crash, harness bug.} Sanitizer fires; LLM judges the cause as a P1/P2 violation in the harness itself.  \emph{Violation} --- back to the build-fix loop.
\item \textbf{No crash, reach miss.} The blob did not drive execution into the target API.  \emph{Violation} --- back to the build-fix loop.
\item \textbf{Crash, real library bug.} Sanitizer fires; LLM judges the cause as a defect in the library (P1--P4 clean on the harness).  \emph{Not a violation} --- submit sanitizer report and harness artifact upstream.
\item \textbf{No crash, reach OK.} The harness handled the attack as intended.  \emph{Not a violation} --- continue.
\end{enumerate}

\emph{Step 2: 600\,s run + coverage unlock.}
Once every blob lands in case~4, QuartetFuzz performs a single 600\,s ASan/LSan run on the binary.
get\_coverage then rebuilds the binary with coverage instrumentation and runs \texttt{llvm-cov}~\cite{llvm} over the 10-minute corpus to produce per-case line and branch coverage numbers.

\emph{Step 3: Submit-eligible probing.}
The agent loops back to Step~1's probing pattern with two new affordances: get\_coverage and submit\_harness.
Submission is gated by a hard constraint and a soft check.
The hard constraint is \emph{both line and branch coverage~$>$~0 and the entry function is dynamically reached};
we use the entry rather than the LG core because Stage~4 has no seed corpus, so core functions deeper in the stack may remain unreached.
The soft check is the LLM's own judgment that the harness is ready, made on the basis of additional AP probes and the coverage numbers.
If the LLM declines to submit, it implicitly signals that some check still fails, and the loop continues with another build-fix iteration.

%% file: sections/implementation.tex
\section{Implementation}
\label{sec:impl}

The system runs on one server with 32 CPU cores and 62\,GB RAM.
Ten workers run in parallel.

We implement QuartetFuzz in Python (\raise.17ex\hbox{$\scriptstyle\sim$}4{,}500
lines of new code) on top of a shared \texttt{BaseAgent} abstraction
that pairs an LLM with an MCP tool server~\cite{mcp}.
Three agents (Logic Group, API Research, Harness Generator) operate over four tool categories: \texttt{code\_view} (source navigation) and \texttt{SAST} (call-graph queries) are shared by all three agents; \texttt{DAST} (build / fuzz / GDB execution) is exposed only to the Harness Generator, which is the only agent that produces and runs binaries; each agent additionally has its own \texttt{terminator} tool that ends its turn loop with the agent's final output.
Table~\ref{tab:agents} (Appendix~\ref{sec:appendix:agents}) details per-agent inputs, outputs, models, turn caps, and how each DAST tool maps to the AP probes from \S\ref{sec:design:probing}.

Each agent runs a turn loop: render prompt, call LLM, dispatch tool calls via FastMCP, repeat.
QuartetFuzz has no built-in static-analysis backend; it consumes a call graph in a JSON schema that any SAST tool can produce. This paper uses Joern~\cite{joern2014} to generate the call graph, with a tree-sitter~\cite{treesitter} fallback for projects Joern does not parse.
Harness build, fuzzing, and coverage collection all use OSS-Fuzz infrastructure; Adversarial Probing (\S\ref{sec:design:probing}) is built on top.

%% file: sections/evaluation.tex
\section{Evaluation}
\label{sec:eval}

We evaluate QuartetFuzz through five research questions:

\begin{itemize}[leftmargin=*]
\item \textbf{RQ1 (Real-World Harness Quality Audit):} Can our
  P1/P2 checks detect and repair quality violations in
  real-world production harnesses?
\item \textbf{RQ2 (Logic Group Quality):} Given a
  natural-language functionality prompt, can the pipeline
  identify entry functions (P3/P4) and produce a
  Logic Group that matches the human-selected target?
\item \textbf{RQ3 (Generation vs.\ Baselines):} How do our
  generated harnesses compare against the human-written gold
  standard and two LLM-based generators (OSS-Fuzz-Gen
  and PromeFuzz)?
\item \textbf{RQ4 (Ablation Study):} What is the contribution of
  each pipeline component---the P2 research stage, the static
  call-graph tools, and the dynamic build/coverage loop?
\item \textbf{RQ5 (Real-World Deployment \& Vulnerability Discovery):} Does the full system
  discover real vulnerabilities in real-world projects?
\end{itemize}

\subsection{Experimental Setup}
\label{sec:eval:setup}

\noindent\textbf{Environment and fuzzing.}
All experiments run on one server (32 cores, 62\,GB RAM) with ten parallel workers.
For RQ5, projects are built and fuzzed inside their OSS-Fuzz Docker images.
Fuzzing uses libFuzzer with ASan and empty seed corpora throughout. LSan is linked at build time and runs by default alongside ASan, so we use ``ASan'' as shorthand for the combined ASan+LSan instrumentation.
Per-harness budget: RQ1 fuzzes each evaluated harness for a single 60\,s run;
RQ3 and RQ4 use $10\times600$\,s and report the per-harness median across the 10 runs;
RQ5 uses a single 10\,h run per harness.

\noindent\textbf{Metrics.}
Following~\cite{klees2018evaluating}, coverage uses LLVM's \texttt{llvm-cov}~\cite{llvm} line and branch (\%).
\emph{Productive rate} (Tables~\ref{tab:rq3_headline}, \ref{tab:ablation}): unlike prior work's build-success rate, we count a harness as failed if it fails to build \emph{or} builds with zero on both coverage metrics---a stricter criterion.
Unproductive harnesses contribute~0 to the cross-case coverage mean (not N/A), so reported averages directly penalise failure rather than hiding it in the denominator.

\noindent\textbf{Models.}
Production configuration is Opus~4.6 for LG/API research and Sonnet~4.6 for harness generation; RQ5 deploys this in full, while RQ1 and RQ3 evaluate harness generation only and use Sonnet~4.6.
RQ2 compares Opus~4.6 vs.\ Gemini~3.1~Pro on the LG task;
RQ4 swaps in Gemini~3.1~Flash to probe model-capability sensitivity.
All production runs use temperature~0; \texttt{max\_tokens} uses each SDK's default. The RQ4 Skill experiment runs Sonnet~4.6, Opus~4.6, Gemini~3.1~Pro, Gemini~3.1~Flash, and GPT-5.4 at each model's default temperature.

\noindent\textbf{Baselines.}
We compare against two LLM-based generators.
OSS-Fuzz-Gen (OFG)~\cite{ossfuzzgen} is Google's production system;
its input format (project + target function) is closest to ours.
PromeFuzz~\cite{promefuzz} (CCS\,'25) is the latest academic system and already directly compared against PromptFuzz~\cite{promptfuzz}, CKGFuzzer~\cite{ckgfuzzer}, OSS-Fuzz-Gen, and hand-written OSS-Fuzz harnesses.
To enable entry-by-entry comparison, we modified both OFG and PromeFuzz so that all three systems share the same I/O, model, and OSS-Fuzz project pinning; the modified code is included in our open-science artifact.
Scheduzz~\cite{scheduzz} and HarnessAgent~\cite{harnessagent} are the systems closest to ours, but Scheduzz is not open-source and HarnessAgent is a concurrent work that is still under review.
We compare their designs and headline numbers in \S\ref{sec:related} and Table~\ref{tab:llm_comparison}.

\noindent\textbf{Dataset.}
We use 100 harnesses from 39 C/C++ OSS-Fuzz projects, drawn from the 586 in RQ1.
Three researchers manually selected cases meeting all of: (i)~the gold harness has stable coverage in OSS-Fuzz Introspector~\cite{introspector}, indicating long-running production exposure on ClusterFuzz~\cite{clusterfuzz}; (ii)~clean build without complex configuration changes; (iii)~$>1\%$ line coverage and no crash on a 600\,s ASan empty-corpus run; (iv)~entry function is a public API.
RQ2--RQ4 use this dataset (full list in Table~\ref{tab:benchmark_full}). During generation, the gold harness is replaced with a stub (\texttt{return~0}) so the agent cannot read it; the original is restored for coverage comparison.

\subsection{RQ1: Real-World Harness Quality Audit}
\label{sec:eval:rq1}

To validate that our P1 and P2 checks assess harness quality, we apply them directly to production harnesses written by experienced developers, reviewed through pull requests, and continuously executed on Google's ClusterFuzz~\cite{clusterfuzz} for years.

We deliberately do not check P3 (security boundary) or P4 (entry-point adequacy) on these human-written harnesses because their authors are the library's own developers and can knowingly cross boundaries or call internal functions, taking responsibility for those choices.
A third-party AI generator has no such authority, so P3/P4 apply only at generation time (RQ2).

\noindent\textbf{Project selection.}
We select 70 C/C++ projects from OSS-Fuzz based on four criteria:
(1)~\emph{impact}: high GitHub star counts, indicating wide
adoption; (2)~\emph{language}: C or C++, where memory-safety
vulnerabilities are most prevalent; (3)~\emph{active maintenance}:
at least one commit within the past month; and
(4)~\emph{accessible issue tracker} (e.g., GitHub/GitLab issues or JIRA): to facilitate upstream reporting.
From these 70 projects we collect 586 production harnesses from OSS-Fuzz (listed in Table~\ref{tab:rq1_projects}).

\noindent\textbf{Methodology.}
The audit reuses Stage~4 (\S\ref{sec:design:stage4}) directly. Only the input changes.
We feed Stage~4 the source plus its built binary as Stage~3's output.
Stage~1 is bypassed because the entry has already been chosen by the original author.
Stage~2 is also bypassed; P2 protocol research instead runs inline at the start of Stage~4, without producing a standalone protocol report.

Stage~4 then proceeds as designed, driven by \emph{Adversarial Probing} (AP, \S\ref{sec:design:probing}).
The agent uses Claude Sonnet~4.6 at temperature~0, with the P1.x/P2.x checklist injected into the prompt; \texttt{static\_analysis} is disabled since the entry is already fixed.
Each repair iteration rewrites the source and re-runs Stage~4's gate.
For RQ1 we fuzz each version (original and repaired) once for 60\,s (default is 600\,s; shortened to keep 586 cases tractable) under ASan/LSan with an empty corpus and compare LibFuzzer edge-coverage counters.
The repair is submitted upstream iff its coverage is \emph{not weaker} than the original's, demonstrating to maintainers that the P1/P2 fix does not regress their existing fuzzing.
If five repair attempts cannot meet this bar, the agent treats its own initial judgment as a false positive and abandons the case.

\noindent\textbf{Results.}
The audit produces two distinct outputs, both legitimate paths through Stage~4's crash triage.

\noindent\emph{(A) Harness-quality violations.}
The P1/P2 review identifies 53 violations (9.0\% of 586) across 28 projects;
the remaining 42 projects are P1/P2-clean, corroborating that our checks are not over-triggered on well-maintained harnesses.
All 53 are submitted upstream. Of these, 45 (85\%) are confirmed, 35 (66\%) fixed or merged, and the rest await review.
One additional P1 report was withdrawn after the maintainer noted the API already handled the edge case. This is our only P1 false positive and harmless to existing fuzzing.

\noindent\emph{(B) Latent library vulnerabilities.}
In two cases, repairing a P2 violation \emph{unmasked} a latent library bug (Table~\ref{tab:vulns_projects}, audit-fix rows).
For openssl, the original harness had a wrong call order.
After we fixed it, the harness triggered a stack-buffer-overread in OpenSSL's DES implementation that had been latent for over 25 years.
For tidy-html5, the original harness was missing a required call.
After we added it, the harness triggered a memory leak.
In both cases, crash triage classified the crash as a real library bug (not a coverage regression) and the bug was fixed upstream.
This unmasking is a direct consequence of P2-driven repair.
Misused harnesses silently mask bugs that hide on the correct usage path.

\noindent\emph{(C) Coverage gain is not the goal.}
Our objective is to fix correctness defects, not to maximize coverage.
Large gains occur when the original is dead code (e.g., opencv/filestorage, libyaml/emitter).
Small gains occur when the harness already exercised the target but had a non-blocking P1/P2 violation (e.g., jq/parse\_stream).
In such cases the coverage figure mainly demonstrates that the fix does not regress existing testing.
Figure~\ref{fig:p1p2_examples} in the appendix shows representative P1 and P2 before/after code diffs.

Table~\ref{tab:confirmed} reports each violation's coverage impact.
We use LibFuzzer edge coverage by default;
three rows use a different metric (cairo/raster and harfbuzz/hb\_set use LLVM line coverage, jq/parse\_stream uses function-execution count).
Notable gains include opencv/filestorage at $+986\%$, libyaml/emitter at $14\times$, and openssl/quic\_server at $+65.7\%$.
14 violations also produced false-positive ASan/LSan crashes that the repair eliminated.
Without the audit these 14 would have surfaced as false-positive bug reports.

\begin{table}[t]
\caption{All 35 fixed (merged upstream) violations with coverage impact;
another 10 are confirmed but not yet merged (Table~\ref{tab:rq1_projects}).
Each row reports a single 60\,s coverage comparison between the repaired harness and the original, empty corpus, ASan/LSan.}
\label{tab:confirmed}
\small
\begin{tabular}{@{}rllll@{}}
\toprule
\textbf{\#} & \textbf{Project} & \textbf{Fuzzer} & \textbf{Princ.} & \textbf{Cov. Impact} \\
\midrule
1  & apache-httpd & parse           & P1    & +11.2\% \\
2  & apache-httpd & uri             & P1    & +2.3\% \\
3  & binutils    & ranlib          & P1    & +11.7\% \\
4  & boost       & filesystem      & P1    & +21.9\% \\
5  & botan       & gcd             & P1    & +29.7\% \\
6  & bzip2       & bzip2\_fd       & P1,P2 & +52.4\% \\
7  & bzip2       & decompress      & P1    & +18.6\% \\
8  & cairo       & raster          & P2    & +14.0\% \\
9  & gdk-pixbuf  & cons            & P1,P2 & fixes FP crash \\
10 & gdk-pixbuf  & file            & P1,P2 & fixes FP crash \\
11 & gdk-pixbuf  & scale           & P1    & fixes FP crash \\
12 & ghostscript & xpswrite        & P1    & +3.1\% \\
13 & harfbuzz    & hb\_set         & P1    & +29$\times$ \\
14 & jq          & parse\_stream   & P1    & +47.0\% \\
15 & lcms        & cgats           & P1    & fixes FP crash \\
16 & lcms        & dict            & P1,P2 & fixes FP crash \\
17 & lcms        & transform\_ext  & P1    & fixes FP crash \\
18 & libarchive  & linkify         & P1    & fixes FP crash \\
19 & libpcap     & rserver         & P2    & +25.0\% \\
20 & libpng      & readapi         & P2    & fixes FP crash \\
21 & libpng      & transforms      & P1,P2 & fixes FP crash \\
22 & libssh2     & ssh2\_client    & P1    & +16.6\% \\
23 & libyaml     & emitter         & P1    & +14$\times$ \\
24 & ndpi        & is\_stun        & P1    & +18.6\% \\
25 & njs         & script          & P1    & fixes FP crash \\
26 & opencv      & filestorage     & P2    & +986\% \\
27 & openssl     & provider        & P2    & +51.4\% \\
28 & openssl     & quic\_server    & P1    & +65.7\% \\
29 & openvpn     & packet\_id      & P1    & fixes FP crash \\
30 & openvpn     & verify\_cert    & P1    & fixes FP crash \\
31 & tidy-html5  & general         & P1,P2 & +33.5\% \\
32 & tidy-html5  & parse\_file     & P1,P2 & +57.7\% \\
33 & unbound     & parse\_packet   & P1    & fixes FP crash \\
34 & wamr        & mutator         & P2    & fixes FP crash \\
35 & zlib        & uncompress3     & P1    & +4.8\% \\
\midrule
\multicolumn{5}{@{}l}{\footnotesize Audit cost: 586 harnesses, $\sim$10\,min each, 10 parallel workers, $\sim$\$720 total.} \\
\bottomrule
\end{tabular}
\end{table}

\subsection{RQ2: Logic Group Quality}
\label{sec:eval:rq2}

\sloppy
To evaluate P3 and P4, we test whether the LG Agent's entry selection matches human expert choices.
Stage~1 (\S\ref{sec:design:stage1}) only emits entries that satisfy the static P3/P4 conditions (public-API reach to core, $\textit{danger}>0$);
RQ2 therefore tests whether the agent's choices among qualifying entries align with human experts'.
If the LG selects the wrong functions, the generated harness fuzzes the wrong code paths regardless of how well P1 and P2 are enforced downstream.

We use the 100-case curated gold dataset (Table~\ref{tab:benchmark_full}) as ground truth.
The dataset labels each harness's \emph{target function}, the API that actually consumes the fuzz-byte input.
A case is a \emph{match} if this target appears in the LG's entry set.
Formally, a gold target $g$ \emph{matches} the LG's entry set $\mathcal{E}$ iff
\[
  \underbrace{\bigl(\exists e \in \mathcal{E}.\;
    \mathrm{bn}(e) = \mathrm{bn}(g)\bigr)}_{\text{direct}}
  \;\;\lor\;\;
  \underbrace{\bigl(\exists e \in \mathcal{E}.\;
    e \in \mathrm{callees}_G(g)\bigr)}_{\text{wrapper}},
\]
where $\mathrm{bn}(\cdot)$ strips namespace prefixes and $\mathrm{callees}_G$ is the single-hop callee set from the project call graph.

For each case we craft a functionality prompt from the fuzzer's upstream pull-request description, documentation, or source code, split into two groups: 60 prompts that name the target function explicitly (\emph{w/ target}), and 40 that contain only a natural-language feature description (\emph{w/o target}).
This contrast separates prompt-driven matching (the target name is given) from code-driven matching (the agent must locate the entry from source alone).

\begin{figure}[h]
\begin{tcolorbox}[colback=blue!5, colframe=blue!40, title=Prompt Example: libyaml/loader]
\small
\textbf{w/ target:}\\[2pt]
``Write a fuzzer for libyaml that targets \texttt{yaml\_parser\_load}. Feed arbitrary bytes as YAML documents to test parser robustness.''\\[6pt]
\textbf{w/o target:}\\[2pt]
``Write a fuzzer for libyaml's YAML document loading functionality. Feed arbitrary malformed YAML input to exercise the parsing and document construction paths.''
\end{tcolorbox}
\caption{Example prompt pair for the same case.}
\label{fig:rq2_prompt_example}
\end{figure}

The pipeline runs a single pass per case. The LG Agent (Claude Opus~4.6) proposes a candidate entry set $\mathcal{E}$ using \texttt{code\_view} and \texttt{static\_analysis} tools, then the Generator writes a first-draft harness.
Neither stage builds or fuzzes the harness.
To compare across model families, we repeat the experiment with Gemini~3.1~Pro using the same pipeline and tools.

\begin{table}[t]
\caption{RQ2: entry-function selection quality on 100 cases.
\emph{w/ target}: prompt names function (60 cases); \emph{w/o target}: feature description only (40 cases).}
\label{tab:rq2_main}
\small
\centering
\begin{tabular}{@{}ll cc@{}}
\toprule
Stage & Prompt & Opus 4.6 & Gemini 3.1 Pro \\
\midrule
LG & w/ target      & 56/60 & 52/60 \\
   & w/o target     & 35/40 & 32/40 \\
   & \textbf{All}   & \textbf{91/100} & \textbf{84/100} \\
\midrule
Harness & w/ target      & 57/60 & 51/60 \\
        & w/o target     & 34/40 & 30/40 \\
        & \textbf{All}   & \textbf{91/100} & \textbf{81/100} \\
\bottomrule
\multicolumn{4}{@{}l}{\footnotesize LG = gold target $\in$ entry set; Harness = gold target called in code.}
\end{tabular}
\end{table}

Table~\ref{tab:rq2_main} reports the results.
With Opus~4.6, the LG stage and the first-draft harness both achieve 91\% match with gold targets.
The \emph{w/o target} group, with zero function names in the prompt, reaches 88\% LG match and 85\% harness match, only 5--10pp below the \emph{w/ target} group.
Entry selection is therefore driven by code structure, not by the prompt content.
Gemini~3.1~Pro achieves 84\% LG match and 81\% harness match, 7--10pp below Opus~4.6, with a wider gap on the harder \emph{w/o target} split.
Both models benefit from the same static-analysis tools; the gap reflects differences in code comprehension ability.

The 9 misses (Opus) span three patterns: (1)~five cases where target names differ across language bindings or require template resolution beyond our static analysis;
(2)~three C++ OO cases where both factory and instance methods are valid entries and the LG agent chose the opposite from gold;
(3)~one case (freerdp) covering 8/9 codec decompressors but missing the rarest.

\fussy

\subsection{RQ3: Generation vs.\ Baselines}
\label{sec:eval:rq3}

\sloppy
RQ3 asks whether QuartetFuzz produces harnesses that match human-written OSS-Fuzz harnesses in coverage.
All systems take the same input (project + target entry function) and run for $10\times600$\,s under LibFuzzer (empty corpus, ASan; median coverage over 10 runs).
All LLM-based systems use Claude Sonnet~4.6 (parameters in \S\ref{sec:eval:setup}) with an identical build-retry budget of~5.

We compare QuartetFuzz against Gold (the human-written OSS-Fuzz harness), OFG~\cite{ossfuzzgen}, and PromeFuzz~\cite{promefuzz}.
Table~\ref{tab:rq3_headline} summarises; per-case data in Table~\ref{tab:full_comparison_multi}.
QuartetFuzz coverage statistically matches Gold.
Across the 100 cases, the per-case difference (QF$-$Gold) averages $+0.03$pp on line and $-0.07$pp on branch.
To check whether this gap is small enough to count as a match, we run a paired TOST equivalence test~\cite{klees2018evaluating} within a $\pm 2$pp tolerance band; it accepts equivalence on both metrics (line $p{=}1.6{\times}10^{-11}$, branch $p{=}2.1{\times}10^{-12}$).
Cliff's $\delta$~\cite{romano2006appropriate} measures the actual size of the gap independent of sample size, and is negligible: $-0.015$ on line, $-0.022$ on branch.

\begin{table}[t]
\caption{RQ3 coverage on 100 cases.
Each harness: $10\times600$\,s LibFuzzer, empty corpus, ASan;
per-harness median across 10 runs;
table is mean over 100 medians.}
\label{tab:rq3_headline}
\small
\centering
\begin{tabular}{@{}lrrrr@{}}
\toprule
 & Gold & OFG & PromeFuzz & QuartetFuzz \\
\midrule
Avg.\ line (\%)      & 17.7 & 10.8 & 12.5 & \textbf{17.7} \\
Avg.\ branch (\%)    & 17.6 &  9.2 & 13.4 & \textbf{17.5} \\
Productive rate      & 100/100 & 64/100 & 74/100 & \textbf{96/100} \\
Avg.\ cost           & --   & \$2.43 & \$1.98 & \$1.65 \\
\bottomrule
\end{tabular}
\end{table}

\noindent\textbf{Per-case breakdown vs.\ Gold.}
The match also holds at the case level: of 100 per-case differences (QF$-$Gold), 72/78 cases (line/branch) fall within $\pm 2$pp, with 17/10 wins and 11/12 losses.
The 17 line wins split into four:
(i) \emph{multi-API drivers} (7 cases) that exercise post-call paths gold's bare-parse driver omits;
(ii) \emph{configuration-varying drivers} (4 cases) that route fuzzer bytes through library options;
(iii) \emph{multi-strategy drivers} (2 cases) that dispatch between distinct init paths from the first byte;
(iv) \emph{initialisation fixes} (4 cases) that add library setup gold's harness omits.

The 11 line losses split into four:
(i) \emph{unproductive harnesses} (4 cases) producing zero line coverage (analysed below);
(ii) \emph{gold's wider input grammar} (3 cases) covering more API breadth than ours;
(iii) \emph{suboptimal driver decisions} (3 cases) where our input handling under-uses the entry;
(iv) one marginal case at the threshold.
Figure~\ref{fig:rq3_winloss} shows one representative win and loss: wabt/wasm2wat (config-varying) and harfbuzz/subset (gold's input grammar).

\noindent\textbf{Comparison with baselines.}
The two baselines fall significantly short.
Per-case ranking across the four systems shows QuartetFuzz finishes first on 42/100 cases and first or second on 73/100, ahead of Gold (29 / 81), OFG (5 / 13), and PromeFuzz (24 / 33).
OFG produces working harnesses for only 64/100 cases; its average coverage trails by 6.9--8.3pp.
PromeFuzz succeeds on 74/100 cases but still trails by 4.1--5.2pp;
even after we extract from its per-project pool the fuzzer calling the gold target (\S\ref{sec:eval:setup}), the lack of per-function steering shows---it occasionally achieves high coverage on trivial targets (e.g., apache-httpd) but underperforms on the rest.

QuartetFuzz reaches a 96/100 productive rate, against 64 for OFG and 74 for PromeFuzz.
This margin reflects the practical value of P1/P2 checks and AP.
Most baseline failures are build errors or zero-coverage harnesses that simple retry cannot fix, but our P1/P2-guided pipeline catches and repairs them before fuzzing begins.
The 4 unproductive cases are agent-introduced harness bugs (wrong include path, wrong API surface, wrong linkage convention, wrong build-include order); all 4 exhaust the 5-attempt build cap in Stage 3 without producing a binary, so none of them reaches Stage 4.

To confirm QuartetFuzz's gap over the two baselines is significant, we run a paired Wilcoxon signed-rank test on the per-case coverage differences.
Both gaps come out significant: vs.\ OFG, $p{<}10^{-11}$ on both metrics with Cliff's $\delta$ medium; vs.\ PromeFuzz, $p{=}1.6{\times}10^{-4}$ on line and $p{=}0.013$ on branch ($\delta$ small).
QuartetFuzz also costs the least at \$1.65/harness, below PromeFuzz (\$1.98) and OFG (\$2.43).
Our P1/P2 checks catch most issues before the build loop, whereas OFG and PromeFuzz exhaust the 5-attempt build cap retrying broken harnesses; OFG still leaves 33/100 cases failing to build.

\noindent\textbf{Adversarial Probing analysis.}
Every one of the 96 productive cases invokes AP at least once, 362 AP calls in total (mean $3.77$, median $3$, max $12$).
22 of these 96 cases need six or more probes, evidence that the agent is constructing attack blobs and reach probes rather than just reading the code.
AP's overall importance is shown by the ablation in \S\ref{sec:eval:rq4}.

\subsection{RQ4: Ablation Study}
\label{sec:eval:rq4}

\begin{figure}[H]
\centering
\includegraphics[width=0.85\columnwidth]{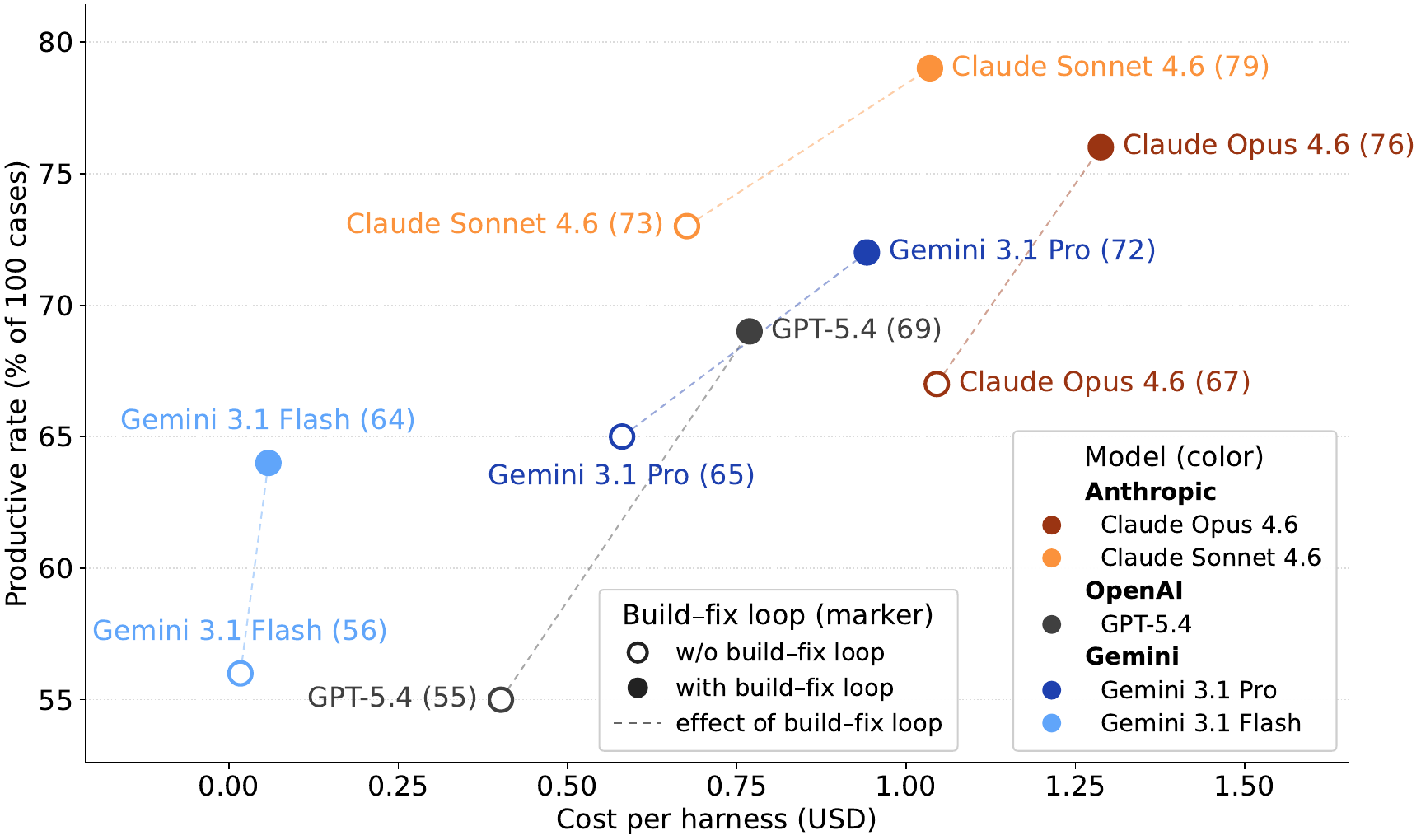}
\caption{Four Principles under a prompt-only Skill-style setup: cost vs.\ productive rate across five models, with and without the build--fix loop.}
\label{fig:rq4_skill}
\end{figure}

We isolate the contribution of each component by disabling it and re-running all 100 cases under the same conditions as RQ3.
Table~\ref{tab:ablation} reports component ablations on Sonnet~4.6;
per-case data is in Table~\ref{tab:ablation_full}.
To test portability, we evaluate an Agent-Skill-style prompt-only configuration on five frontier models (Figure~\ref{fig:rq4_skill}).

\begin{table}[t]
\caption{RQ4 component ablations on the 100-case dataset, run identically to Table~\ref{tab:rq3_headline}.
\emph{Full} reproduces the QuartetFuzz row from Table~\ref{tab:rq3_headline}; the remaining columns each ablate one piece: \emph{w/o AP} disables the adversarial probe gate; \emph{w/o P2} removes the API protocol report; \emph{w/o Static} removes call-graph tools; \emph{Flash} swaps the generator model to Gemini 3.1 Flash.}
\label{tab:ablation}
\small
\centering
\begin{tabular}{@{}l rrrrr@{}}
\toprule
 & \textbf{Full} & w/o AP & w/o P2 & w/o Static & Flash \\
\midrule
Avg.\ line (\%)     & \textbf{17.7}    & 16.9   & 16.2   & 17.5   & 17.5 \\
Avg.\ branch (\%)   & \textbf{17.5}    & 16.6   & 15.9   & 17.3   & 17.3 \\
Productive rate     & \textbf{96/100}  & 84/100 & 83/100 & 88/100 & 89/100 \\
Avg.\ cost          & \textbf{\$1.65}  & \$1.04 & \$1.35 & \$1.14 & \$1.64 \\
\bottomrule
\end{tabular}
\end{table}

\noindent\textbf{Without Adversarial Probing.}
Productive rate drops 96$\to$84 ($-12$pp);
line $-0.8$pp, branch $-0.9$pp;
cost falls \$0.61/case (the AP overhead).
The 12 lost cases all compile and run but never reach the target---without AP, no signal catches this and they slip past the build--fix loop.

\noindent\textbf{Without the P2 report.}
Removing the API-protocol report drops productive rate from 96 to 83 ($-13$pp) and line coverage by 1.5pp.
Cost falls modestly to \$1.35 because the saved P2 stage tokens are partly offset by the agent burning extra turns rediscovering protocol details on its own.
The net effect is the largest quality regression of any single ablation.

\noindent\textbf{Without static analysis.}
Removing call-graph tools drops productive rate from 96 to 88 ($-8$pp) and line coverage by 0.2pp ($-0.2$pp branch).
Cost falls to \$1.14 because the agent has fewer tools to invoke;
the LG and harness agents recover most structural context from \texttt{code\_view} alone, so the pipeline degrades gracefully---static analysis is the least critical of the three components.

\noindent\textbf{Model swap (Gemini 3.1 Flash).}
Swapping in Gemini 3.1 Flash gives comparable coverage ($-0.2$pp line) but productive rate drops to 89 and cost is \$1.64 (essentially unchanged from Full at \$1.65);
the pipeline still recovers 89 working harnesses on a weaker model, demonstrating that the principles transfer across model families even when raw capability differs.

\noindent\textbf{Skill-style portability.}
Inside the same pipeline, we keep only \texttt{code\_view} and a prompt-only P1--P4 checklist (with and without the build--fix loop), approximating an Anthropic Agent Skill~\cite{anthropic-skills} setup on five frontier models.
The build--fix loop helps every model (+6--14pp; Figure~\ref{fig:rq4_skill});
stronger models do not substitute for it (Sonnet 4.6: $73\!\to\!79$;
Opus 4.6: $67\!\to\!76$).
The best Skill-style setup reaches 79/100, 17pp below the full pipeline's 96, quantifying the infrastructure contribution.

\subsection{RQ5: Real-World Deployment \& Vulnerability Discovery}
\label{sec:eval:rq5}

Fuzz harnesses exist to find bugs~\cite{aixcc}.
We therefore deploy QuartetFuzz on 23 real-world open-source projects (C/C++, Java, JavaScript) to evaluate the full pipeline and its vulnerability-discovery ability.

\noindent\textbf{Vulnerability Discovery.}
The 23 projects come from two channels: (i)~12 projects (10~C/C++, 1~Java, 1~JavaScript; bold in Table~\ref{tab:vulns_projects}) where we established direct collaboration (Figure~\ref{fig:developer_survey}) with maintainers or were asked by our sponsor, and (ii)~11 additional projects following the same selection criteria as RQ1.
The full project list was fixed before deployment and collaborations were established before any harness was generated and run;
no project was selected post-hoc based on whether it produced a crash.
ghidra and graaljs are not in OSS-Fuzz upstream; we wrote our own harnesses and build scripts.

\begin{figure}[t]
\centering
\includegraphics[width=\columnwidth]{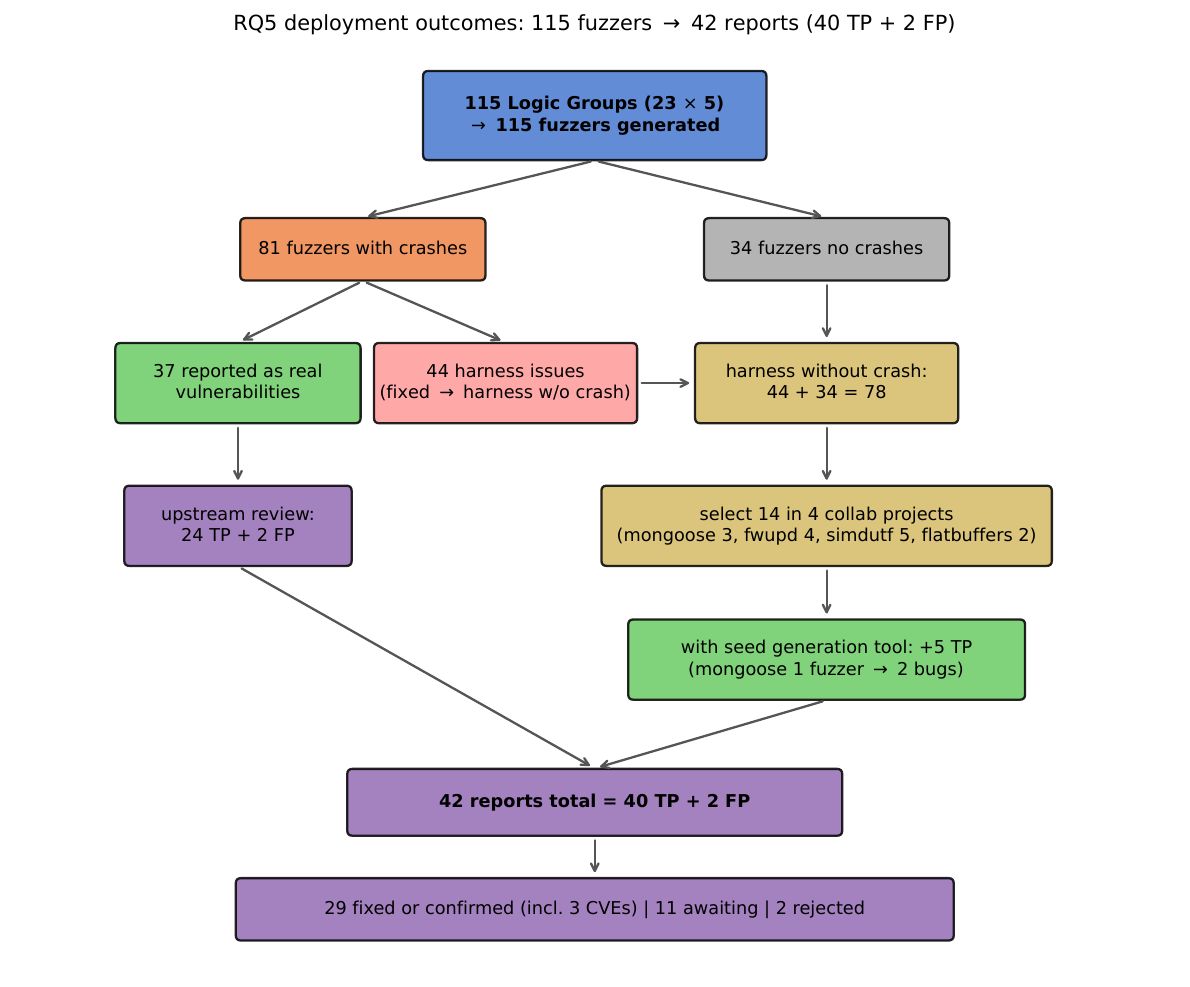}
\caption{RQ5 deployment funnel: 115 fuzzers (23$\times$5 LGs) $\to$ 42 reports (40 TP + 2 FP) $\to$ 29 fixed or confirmed.}
\label{fig:rq5_funnel}
\end{figure}

Figure~\ref{fig:rq5_funnel} shows the full deployment funnel.
The LG agent generates 5--10 Logic Groups (LGs) per project and keeps the top-5 by danger score, yielding 115 LGs.
These 115 LGs feed full-pipeline harness generation.
Adversarial Probing and the Stage~4 600\,s gate test run together produced 81 sanitizer crashes.
Crash triage classified 44 as harness-side P1/P2 violations (fixed in place, not reported) and 37 as real library bugs reported upstream.
On 4 collaboration projects, we re-ran 14 fuzzers with no crash (mongoose 3, fwupd 4, simdutf 5, flatbuffers 2) using a targeted seed generator~\cite{seedgen}, finding 5 additional real bugs (mongoose's HTTP fuzzer alone yielded 2).
Total LLM cost: approximately \$300 for the 115-LG full pipeline and \$500 for seed-generation reruns across the 4 collaboration projects ($\sim$\$100/project).

\begin{figure}[t]
\centering
\begin{subfigure}{0.34\columnwidth}
  \centering
  \includegraphics[width=\linewidth]{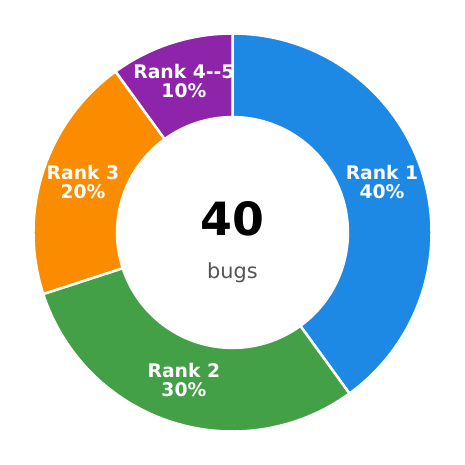}
  \caption{LG rank.}
  \label{fig:rank_dist}
\end{subfigure}\hspace{0.02\columnwidth}%
\begin{subfigure}{0.63\columnwidth}
  \centering
  \includegraphics[width=\linewidth]{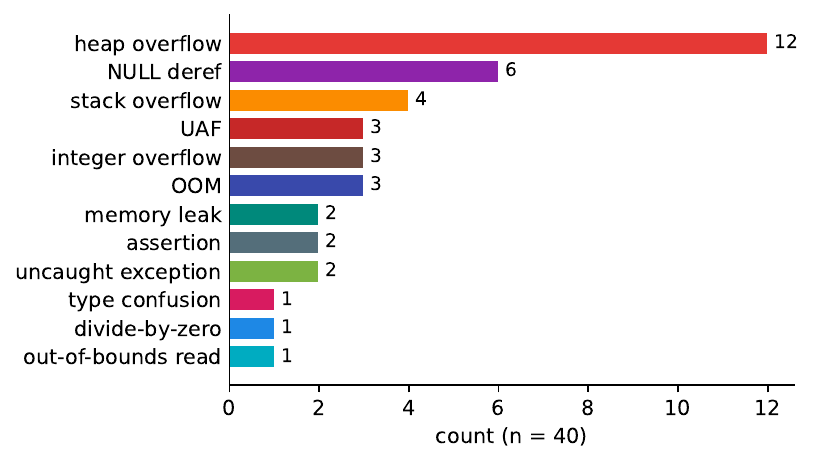}
  \caption{Vulnerability types.}
  \label{fig:vuln_types}
\end{subfigure}
\caption{Distribution of the 40 real bugs (2 FPs excluded): 40\% at rank~1, 70\% in top~2.}
\label{fig:rq5_dist}
\end{figure}

In total QuartetFuzz submitted \textbf{42 reports} (40 real + 2 FPs):
20 fixed, 9 confirmed (including 3 CVEs), 11 awaiting review, 2 rejected---a \textbf{4.8\% FP rate}, the lowest among LLM-based generators (Table~\ref{tab:fp_comparison}).
Of the 42 reports, 24 are from collaboration projects (16 fixed, 6 confirmed, 2 FP) and 18 from random projects (4 fixed, 3 confirmed, 11 awaiting);
per-project bug-finding is comparable (2.0 vs.\ 1.6 reports/project), and the higher acknowledge rate on collaboration projects is mostly maintainer latency.
All collaboration-project bugs were independently reviewed by our sponsor and earned a \$120{,}000 bounty.
Figure~\ref{fig:rank_dist} shows the LG-rank distribution and Figure~\ref{fig:vuln_types} the vulnerability-type distribution of the 40 real bugs;
the rank distribution concentrates at the top, confirming that the LG ranking prioritises vulnerability-prone targets.

\noindent\textbf{Cross-language deployment.}
With LLM-only LG generation and ranking, PDFBox (Java) yielded 3 vulnerabilities (all fixed) and GraalJS (JavaScript) yielded 2 (one fixed, one confirmed)---all 5 without false positives.

\noindent\textbf{Controlled validation of LG ranking.}
Since we only deployed rank-1--5, ``100\% in rank~1--5'' is true by construction.
On the 5 projects where rank-1--5 found bugs and the rank comes from a real danger score (not Static-Analysis Fallback (SAF); see Table~\ref{tab:vulns_projects})---mongoose, fwupd, libwebp, opc-ua, rapidjson---we deploy 25 additional rank-6--10 harnesses under the same 10\,h budget and dedupe (ASan top-3 frame).
Ranks 6--10 yield \textbf{zero} new memory-safety crashes (only 2 resource-exhaustion artifacts; Table~\ref{tab:c2_saturation}, Appendix~\ref{sec:appendix:saturation})---the ranking saturates the practically findable bug set.

\noindent\textbf{Upstream adoption.}
Beyond accepting bug reports, 4~maintainer teams produced new fuzzing artifacts from our work: OpenSSL adopted our quic-server rewrite (2.64$\times$ branch coverage);
fwupd generalized our PoC into a generic \texttt{FuDevice} fuzzing interface;
pdfbox committed our crash inputs as JUnit regression tests;
libwebp enhanced their own fuzzer using ours as reference.

%% file: sections/discussion.tex
\section{Discussion}
\label{sec:discussion}

\noindent\textbf{Mathematical specifications vs.\ practical implementation.}
In the abstract we defined P1--P4 as mathematical specifications.
However, the actual implementation does not match the mathematical definitions because the mathematical predicates are not decidable over source code alone.
We approximate them with syntactic sub-checks (P1.1--P1.8, P2.1--P2.8; Table~\ref{tab:probe_attribution}), AP probes, and call-graph reachability with an LLM fallback.
We are fully aware this gap reduces certainty in our claims and could place the paper in a vulnerable position.
We chose to publish anyway because, as \S\ref{sec:intro} documents, harness quality has grown into a serious engineering problem as automated harness generation scales, and to our knowledge no prior work has defined what \emph{correctness} means for an LLM-generated harness.
Our framework offers reasonable approximations grounded in actual audit and generation practice, with every check tied to a concrete artifact (source line, call-graph edge, sanitizer fire).
We treat it as a starting definition, not a final proof, and hope future work builds on this foundation.

\noindent\textbf{Metric limitations.}
Prior automated harness-generation work evaluates systems with line coverage, branch coverage, and build success rate.
These remain necessary but are no longer sufficient.
Producing a harness that compiles and reaches some coverage is now within easy reach of an LLM agent.
In the LLM era the bar should move from ``did it build and run'' to ``is the harness correct''.
Throughout this work, our target is to make the system produce harnesses an experienced developer would have written, not ones that maximise coverage on a benchmark.
We propose P1--P4 as that next-level criterion.
Because we do not yet have a fully deterministic checker, we did not use P1--P4 as the primary evaluation metric in RQ3 and RQ4.
But this is exactly why RQ1, RQ2, and RQ5 matter.
RQ1 validates P1/P2 on production harnesses with experienced developer review, RQ2 validates P3/P4 against human-selected targets, and RQ5 validates the full system in real-world deployment.

\noindent\textbf{Generated vs.\ human-written harnesses.}
P1 and P2 apply to both.
A leak or misused API is a bug whoever wrote the harness; we audit human harnesses against both in \S\ref{sec:eval:rq1}.
P3 and P4 do not.
A library maintainer can knowingly cross a boundary or call an internal helper;
an LLM that has never touched the project cannot.
We therefore enforce P3 and P4 only on our generator.
P3 is a preference, not a hard constraint: when no public entry reaches the target, the generator falls back to the most direct internal entry that preserves as many boundaries as possible.

\noindent\textbf{Seeds.}
RQ1--RQ4 use empty corpora throughout, and RQ5 also defaults to empty corpora.
In practice seeds matter, since an empty corpus cannot produce a valid HTTP request or PNG.
On 4 of the 23 RQ5 projects we additionally paired harnesses with a targeted seed generator~\cite{seedgen}, which yielded 5 of the 40 bugs.

\noindent\textbf{Limitations and future work.}
(1)~Our pipeline depends on OSS-Fuzz, LibFuzzer, ASan/LSan, and C/C++;
pdfbox (Java), graaljs (JS), and non-OSS-Fuzz collaboration projects required pipeline and architecture adjustments.
Future work will extend to mainstream fuzzers (AFL++, HonggFuzz) and other languages (Rust, Go, Python).
(2)~RQ3 and RQ4 use line/branch coverage and productive rate.
The LLM era calls for stronger metrics such as core-function or core-file coverage, which combined with P1--P4 and our 100-case dataset can form a deterministic benchmark for LLM-generated harnesses.
(3)~All experiments use empty corpora;
high-quality per-case seeds that drive execution into the core function would surface deeper bugs.
(4)~Joern fails on some projects (heavy C++ templates, vtable indirection) and the tree-sitter fallback gives a basic but lower-precision call graph.
Stronger pointer/alias analysis (e.g., SVF~\cite{svf2016}) would further improve P3/P4 accuracy.

%% file: sections/related_work.tex
\section{Related Work}
\label{sec:related}

\noindent\textbf{Traditional harness generation.}
Consumer-based approaches extract API usage from existing code: FuzzGen~\cite{fuzzgen}, FUDGE~\cite{fudge}, WINNIE~\cite{winnie}, APICraft~\cite{apicraft}, libErator~\cite{liberator}, and WildSync~\cite{wildsync}.
They depend on consumer-code availability.
Constraint-based approaches enforce API protocols at runtime: Hopper~\cite{hopper} (93.52\% API coverage but 51\% spurious crashes), AFGen~\cite{afgen} (precision 46.97\%$\to$94.55\% via constraint tracing), NEXZZER~\cite{nexzzer} (filters 93.96\% of crashes as API misuse), GraphFuzz~\cite{graphfuzz}, DAISY~\cite{daisy}, and Rubick~\cite{rubick}.
SyzGen~\cite{syzgen} targets the analogous problem for OS syscalls.
None defines correctness a priori at the source level.

\noindent\textbf{LLM-based harness generation.}
PromptFuzz~\cite{promptfuzz} treats generation as a fuzzing loop over prompts ($1.61\times$ branch coverage over OSS-Fuzz on 14 libraries).
CKGFuzzer~\cite{ckgfuzzer} uses a CodeQL~\cite{codeql}+Tree-sitter~\cite{treesitter} knowledge graph;
84.4\% of its 199 crashes stem from API misuse.
PromeFuzz~\cite{promefuzz} builds structured knowledge bases over AST metadata, API docs, and call sequences;
a DeepSeek-R1 sanitizer raises precision from 2.7\% to 89.7\%.
OSS-Fuzz-Gen~\cite{ossfuzzgen} integrates LLM generation into Google's pipeline with Fuzz Introspector and found CVE-2024-9143 in OpenSSL.
HarnessAgent~\cite{harnessagent} is a tool-augmented agent with LSP~\cite{lspspec} and Tree-sitter~\cite{treesitter};
reports 87\%/81\% three-round success but no re-runnable artifact.
Scheduzz~\cite{scheduzz} is the only prior tool applying constraints before generation (Prolog over type/LLM-extracted API dependencies);
no public artifact.
LLM4FDG~\cite{llm4fdg} reports a 34\% API-misuse rate;
FDFactory~\cite{fdfactory}, TitanFuzz~\cite{titanfuzz} extend to deep-learning libraries;
position work on reliable LLM-driven fuzzing~\cite{reliable} reaches a similar conclusion.
Every prior tool relies on behavioural proxies rather than source-level correctness checked before fuzzing (Table~\ref{tab:llm_comparison}).
Adjacent LLM-for-security work~\cite{iris,llm4vuln,kernelgpt,codebert,vuldeepecker,llm4secsurvey} targets vulnerability detection rather than harness synthesis.

\noindent\textbf{Harness quality assessment.}
To our knowledge, no prior work audits production harnesses systematically or defines correctness a priori.
AFGen's Constraints Tracer~\cite{afgen} checks API constraints post-crash;
OGHarn's three oracles~\cite{ogharn} verify compilation, execution, and coverage;
deepSURF~\cite{deepsurf} treats true positives as crashes from contract-respecting use, mapping to our P2.
We define quality as source-level conditions checked before fuzzing and validate on 586 production harnesses (45 confirmed, 35 fixed).
We also release 100 annotated harnesses as the first labeled dataset for harness quality.

%% file: sections/conclusion.tex
\section{Conclusion}
\label{sec:conclusion}

We define the \emph{Four Principles}, four source-level conditions (P1--P4) that a harness must satisfy before fuzzing begins, and build \emph{QuartetFuzz}, an LLM-agent generator that enforces them through a four-stage pipeline of Logic Group selection, API protocol research, static-driven build, and adversarial validation.
Our audit of 586 production OSS-Fuzz harnesses flagged 53 violations (45 confirmed, 35 fixed) and exposed 2 long-latent library bugs.
QuartetFuzz matched human gold coverage on 100 cases (TOST $\pm 2$pp, $p{<}10^{-10}$), beat OSS-Fuzz-Gen by 6.9--8.3pp and PromeFuzz by 4.1--5.2pp, and shipped 42 bug reports across 23 projects with a 4.8\% FP rate (29 fixed or confirmed, 3 CVEs).
Across audit and generation, the same P1/P2 checks intercepted 58 harness-induced crashes that would have been false-positive reports.
Harness quality is the binding constraint on fuzzing effectiveness, and the practical way to enforce it is to bake source-level checks into the generator before fuzzing begins.

%% file: main.bbl

%% file: sections/ccs_meta.tex
\section{Open Science}
\label{sec:openscience}

We release the artifacts that support the paper's claims through three repositories covering the system, the dataset, and the two re-runnable baselines:

\begin{itemize}[leftmargin=*]
\item \textbf{QuartetFuzz: system + dataset.}
LG agent, API Research agent, Harness Generator, and P1--P4 checkers;
the 100-case gold-standard dataset with P1--P4 labels (Table~\ref{tab:benchmark_full});
and \texttt{run\_subset.sh} reproducing RQ3 and RQ4 end-to-end.
\url{https://github.com/OwenSanzas/QuartetFuzz}

\item \textbf{Modified PromeFuzz baseline.}
Our shared-I/O patch over PromeFuzz~\cite{promefuzz} that aligns model, OSS-Fuzz project pinning, and per-target input handling with QuartetFuzz, plus a matching run script.
\url{https://github.com/OwenSanzas/PromeFuzz}

\item \textbf{Modified OSS-Fuzz-Gen baseline.}
Our shared-I/O patch over OSS-Fuzz-Gen~\cite{ossfuzzgen};
same alignment, README, and run script.
\url{https://github.com/OwenSanzas/oss-fuzz-gen}
\end{itemize}

\noindent\textbf{Subset and embargo.}
The full live-rerun artefact ($10\times600$\,s traces, OSS-Fuzz images) exceeds \textbf{230\,GB} and is infeasible to host on GitHub; \texttt{run\_subset.sh} targets 25 lightweight cases chosen as a representative RQ3 slice (runnable in $\sim$30\,min).
We retain the right to re-run the full 100 on request.
Withheld until camera-ready: actual CVE numbers (anonymised in Table~\ref{tab:vulns_projects}), 11 embargoed RQ5 reports, and PoCs / sanitizer traces for unpatched vulnerabilities.

\section{Ethical Considerations}
\label{sec:ethics}

\noindent\textbf{Scope.}
42 RQ5 reports across 23 projects (29 fixed/confirmed with 3 CVEs, 11 awaiting, 2 FPs) plus 53 P1/P2 violations from the 70-project RQ1 audit (45 confirmed, 35 merged).
12 of 23 RQ5 projects are collaborations with prior maintainer or sponsor consent;
the remaining 11 plus the 70 RQ1 projects are in OSS-Fuzz and receive submissions as standard PRs or advisories.
All targets public; no human subjects, no user data; the disclosure protocol below was established before any audit or generation run.

\noindent\textbf{Stakeholders.}
Maintainers receive reproducible PoCs and root causes;
end users gain patched dependencies via the 42 RQ5 fixes;
the 35 P1/P2 fixes repair production harnesses already deployed but silently broken;
the community gains a 100-case labelled benchmark.
Affected parties---users of unpatched libraries (mitigated by embargo and CVE anonymisation) and uncontacted maintainers in 28 audit projects (mitigated by submitting only fixes that can be declined)---face minimal residual exposure.
Adversary prerequisites (frontier-model access, source, OSS-Fuzz format) are project-owner-side, favouring defensive use.

\noindent\textbf{Risks and mitigations.}
Net defensive value (64 upstream fixes, 58 harness-induced false positives blocked pre-triage) against three risks: information leak before patch, dual-use uplift, maintainer burden.
Every report uses the maintainer's designated channel with root-cause and fix, held until patched; CVE numbers and embargoed reports are anonymised in Table~\ref{tab:vulns_projects}.
Released artefacts are limited to the system, the 100-case benchmark, and baseline scripts---no exploit catalog.
Collaboration bugs were independently re-validated by our anonymised sponsor before bounty payout, on the same protocol as unsponsored projects.

\noindent\textbf{Limitations.}
No external ethics consultation;
our stakeholder list may omit downstream commercial users and supply-chain organisations.
Long-term effects on OSS maintainer practice---whether sustained automated quality auditing shifts the review burden onto maintainers---are out of scope.
We will incorporate reviewer feedback into the camera-ready.

\section{Generative AI Usage}
\label{sec:genai}

\noindent\textbf{System implementation.}
${\sim}90\%$ of the framework (agent loop, MCP tools, build / runner / coverage glue, evaluation scripts) was written and maintained by the authors;
Claude (Opus / Sonnet 4.6) generated only auxiliary workflows (small data-processing snippets, plotting scripts), each admitted only after standard peer code review.

\noindent\textbf{Writing assistance.}
The authors wrote paper structure and prose by hand; Claude (Opus 4.6) polished the manuscript, reviewed cross-section consistency, and assisted with figures and tables.
Every Claude-suggested edit was author-reviewed before commit.

\balance

%% file: sections/appendix.tex
\appendix

\begin{table*}[!t]
\caption{QuartetFuzz MCP agents (top) and the tools each category exposes (bottom).}
\label{tab:agents}
\footnotesize
\centering
\setlength{\tabcolsep}{8pt}

\begin{tabular}{@{}llllr@{}}
\toprule
\textbf{Agent} & \textbf{Input $\to$ Output} & \textbf{Model} & \textbf{Tool categories} & \textbf{Turn cap} \\
\midrule
Logic Group       & source repo $\to$ top-5 LGs                & Opus 4.6   & code\_view, SAST, terminator        & 50 \\
API Research      & one LG $\to$ P2 protocol report            & Opus 4.6   & code\_view, SAST, terminator        & 30 \\
Harness Generator & LG + P2 report $\to$ AP-passing binary     & Sonnet 4.6 & code\_view, SAST, DAST, terminator  & 50 \\
\bottomrule
\end{tabular}

\vspace{8pt}

\begin{tabular}{@{}p{2.0cm}p{8.5cm}p{4.5cm}@{}}
\toprule
\textbf{Category} & \textbf{Tools} & \textbf{Role} \\
\midrule
code\_view & read\_file, list\_directory, search\_files, list\_existing\_fuzzers                    & source navigation, available to all agents \\
\midrule
SAST tool & get\_callers, get\_callees, find\_definition, forward\_reach, reverse\_reach, public\_entries\_for, public\_entries\_for\_batch, is\_public\_api              & call-graph queries (forward + reverse, with public-API filtering for the core-first LG search), available to all agents \\
\midrule
\multirow{3}{*}[-1.2em]{DAST tool} & build\_harness, get\_coverage                                                       & build harness in OSS-Fuzz Docker; coverage measurement \\
\cmidrule(lr){2-3}
        & AP\_run\_check                                                                            & \textit{AP Probe~2} (\S\ref{sec:design:probing}); P1.x dynamic check \\
\cmidrule(lr){2-3}
        & AP\_reach\_check                                                                          & \textit{AP Probe~1} (\S\ref{sec:design:probing}); P2.x dynamic check \\
\midrule
terminator & submit\_logic\_group / submit\_p2\_report / submit\_harness                              & per-agent terminator tools (one each, ends the agent's turn loop with its final output) \\
\bottomrule
\end{tabular}
\end{table*}

\section{Controlled Saturation of LG Ranking}
\label{sec:appendix:saturation}

Both rank-6--10 OOMs pass our P1.5 buffer-safety check (libwebp harness caps fuzz input at $2^{22}$\,B, rapidjson at $2^{20}$\,B);
the out-of-memory states originate from unbounded library-internal allocation under bounded fuzz input, classified as resource-exhaustion artifacts rather than memory-safety crashes.
Both have been reported to the respective upstreams.

\begin{table}[H]
\caption{LG-ranking saturation on 5 projects: rank-1--5 (deployed) vs rank-6--10, matched 10\,h LibFuzzer/ASan budget.
Bug counts deduplicated by ASan top-3 frame.}
\label{tab:c2_saturation}
\centering
\footnotesize
\begin{tabular}{@{}lrr@{}}
\toprule
\textbf{Project} & \textbf{Rank 1--5 bugs} & \textbf{Rank 6--10 unique new} \\
\midrule
mongoose   & 3 & 0 \\
fwupd      & 2 & 0 \\
libwebp    & 2 & 1 (OOM) \\
opc-ua     & 2 & 0 \\
rapidjson  & 1 & 1 (OOM) \\
\midrule
\textbf{Total} & \textbf{10} & \textbf{2} (resource exhaustion only) \\
\bottomrule
\end{tabular}
\end{table}

\begin{table}[H]
\caption{False-positive rate comparison.
Crash and FP-rate values for prior tools are taken verbatim from the cited papers;
the QuartetFuzz row reflects the 42 reports submitted in this work and the 2 maintainer-rejected cases (Table~\ref{tab:vulns_projects}).}
\label{tab:fp_comparison}
\small
\centering
\begin{tabular}{lrr}
\toprule
Tool & Crashes & FP Rate \\
\midrule
NEXZZER~\cite{nexzzer} & 7,291 & 93.96\% \\
UTop\'ia~\cite{utopia} & 1,167 & 41.6\% \\
Hopper~\cite{hopper} & 51 & 51.0\% \\
AFGen (no constraints)~\cite{afgen} & 660 & 53.03\% \\
AFGen (full)~\cite{afgen} & 660 & 5.45\% \\
PromptFuzz~\cite{promptfuzz} & 44 & 13.6\% \\
PromeFuzz~\cite{promefuzz} & 29 & 10.3\% \\
\textbf{QuartetFuzz} & \textbf{42} & \textbf{4.8\%} \\
\bottomrule
\end{tabular}
\end{table}

\section{Danger Score Sensitivity}
\label{sec:appendix:danger_sensitivity}

The danger score (Eq.~\ref{eq:danger}) traverses the call graph up to depth $D$, default $D{=}20$.
SAF projects rank LGs by LLM judgment rather than the formula, so $D$ does not apply to them;
we sweep $D \in \{10, 15, 20, 25, 30\}$ on all RQ5 projects whose deployment ranking is formula-driven (SAF projects are excluded because indirect dispatch or unsupported source language prevents the call graph from producing a meaningful danger value).

No project changes its rank-1 LG or top-3 LG set at any $D$ vs $D{=}20$;
per-project Spearman $\rho$ is $1.000$ for $D \in \{15, 25, 30\}$ and $\ge 0.9$ for $D{=}10$ (mean $0.994$), with the only deviation a single adjacent swap at ranks 4--5 in one project.
$D{=}20$ sits on a stable plateau in $[10, 30]$;
we adopt it as a conservative default that captures transitive reach in deeper call graphs without affecting top selections in shallower ones.

\section{MCP Agent and Tools}
\label{sec:appendix:agents}

Table~\ref{tab:agents} is keyed to \S\ref{sec:impl}.
The two AP probes in the DAST row are already specified in \S\ref{sec:design:probing} (Figure~\ref{fig:ap_modes}).
The SAST row uses semantic names whose intent may not be obvious at a glance;
Figure~\ref{fig:sast_tools} fixes the input/output for each.

\begin{figure}[H]
\begin{tcolorbox}[colback=gray!8, colframe=gray!60,
  title=SAST tools: input $\to$ output]
\small
\textbf{find\_definition}\\
\quad function name\\
\quad $\to$ definition site(s): file, line, language, external flag, complexity\\[3pt]
\textbf{forward\_reach}\\
\quad function $f$, depth $D$ (default 20)\\
\quad $\to$ functions reachable from $f$ within $D$ hops\\[3pt]
\textbf{reverse\_reach}\\
\quad function $f$, depth $D$\\
\quad $\to$ functions that reach $f$ within $D$ hops, each tagged\\
\quad\phantom{$\to$ }\textsc{public} / \textsc{internal} / \textsc{unknown}\\[3pt]
\textbf{public\_entries\_for}\\
\quad an internal core function\\
\quad $\to$ public APIs from which the core is reachable\\
\quad\phantom{$\to$ }(candidate harness entries for that core)\\[3pt]
\textbf{public\_entries\_for\_batch}\\
\quad list of core functions\\
\quad $\to$ per-core public APIs (in one tool call)\\[3pt]
\textbf{is\_public\_api}\\
\quad function name\\
\quad $\to$ \textsc{public} (defined under \texttt{include/}, \texttt{public/}, \texttt{api/})\\
\quad\phantom{$\to$ }/\,\textsc{internal} (under \texttt{internal/}, \texttt{private/}, \texttt{src/},\\
\quad\phantom{$\to$ \,/\,}\texttt{core/}, \texttt{util/}, \texttt{common/}, \texttt{impl/}, \texttt{detail/})\\
\quad\phantom{$\to$ }/\,\textsc{unknown} (path matches neither; agent reads\\
\quad\phantom{$\to$ \,/\,}the header to decide)
\end{tcolorbox}
\caption{Input/output of the six non-trivial SAST tools.}
\label{fig:sast_tools}
\end{figure}

\clearpage

\begin{table*}[t]
\caption{\textbf{The four principles: per-sub-check definitions and verification.}
P1 and P2 each have eight sub-checks; P3 cascades through three (structural $\to$ LLM-fallback $\to$ LLM-only when no call graph); P4 has two (structural $+$ LLM-only when no call graph).
Of the 16 P1/P2 sub-checks, 12 are operationalised by Adversarial Probing (AP) and 4 by static review.}
\label{tab:probe_attribution}
\scriptsize
\centering
\setlength{\tabcolsep}{4pt}
\begin{tabular}{@{}p{0.7cm}p{2.2cm}p{6cm}p{2.4cm}p{1.4cm}p{2.7cm}@{}}
\toprule
ID & Name & Purpose \emph{(it checks)} & Probe input & Oracle/Tool & Fail signal \\
\midrule
\multicolumn{6}{@{}l}{\textit{\textbf{Principle 1: Logic Correctness} --- 8 sub-checks on the harness source}} \\
\midrule
P1.1 & Resource leaks            & release of every alloc / fd / lock / handle on every exit path                           & error-path input               & LSan         & Sanitizer crash report \\
P1.2 & Use-after-free            & no pointer read or pass after free                                                       & free-then-use input            & ASan         & Sanitizer crash report \\
P1.3 & Stale state               & no static or global state across fuzz iterations                                         & two distinct inputs in seq     & ASan/LSan    & Sanitizer crash report${}^{*}$ \\
P1.4 & Input flow                & fuzz-byte flow to target API (call not driven by constants)                              & arbitrary input + marker bytes & GDB          & API call site silent \\
P1.5 & Buffer safety             & bounded fuzz-buffer reads; null-terminated C strings; length-checked indexing            & oversized input                & ASan         & Sanitizer crash report \\
P1.6 & Size checks               & early-return when input is below API minimum                                             & undersized input               & GDB          & API breakpoint fires \\
P1.7 & Undefined behaviour       & no out-of-bounds / null-deref / signed-overflow / unaligned-cast                         & ---                            & LLM semantic${}^{\dagger}$ & flags signed-overflow / unaligned-cast / null-deref pattern \\
P1.8 & No reimplementation       & real library code, not a local stub or copy                                              & ---                            & LLM semantic & LLM detects local copy \\
\midrule
\multicolumn{6}{@{}l}{\textit{\textbf{Principle 2: API Protocol Compliance} --- 8 sub-checks on the harness--library interface}} \\
\midrule
P2.1 & Init sequence             & required predecessors in order before target call                                        & arbitrary input                & GDB          & init breakpoint silent before target \\
P2.2 & Parameter construction    & right type / range / owner / lifetime per parameter                                      & invalid parameters             & ASan         & Sanitizer crash report \\
P2.3 & Object lifecycle          & create $\to$ configure $\to$ use $\to$ destroy lifecycle on opaque objects               & arbitrary input                & GDB          & any of create/use/destroy missing \\
P2.4 & Return value handling     & return-code check; error-branch run; gated output use                                    & error-return input             & GDB          & error branch silent \\
P2.5 & Cleanup sequence          & release of every resource on all exit paths in API order                                 & early-exit input               & LSan         & Sanitizer crash report \\
P2.6 & API existence             & every called function exported by the library at pinned build                            & ---                            & static call graph & not under public-header path \\
P2.7 & Co-call constraints       & paired APIs co-occurring; mutex APIs not                                                 & ---                            & LLM semantic & LLM detects mismatched pair \\
P2.8 & Prerequisite state        & external state (fds, sockets, env, threads) set up before and torn down after            & arbitrary input                & GDB          & prerequisite silent before target \\
\midrule
\multicolumn{6}{@{}l}{\textit{\textbf{Principle 3: Security Boundary Respect} --- public vs.\ internal entry boundary, three cascading checks}} \\
\midrule
P3.1 & Boundary respect          & public-API reach to target core (call graph)                                            & ---                            & static call graph & $E_{\text{pub}} = \emptyset$ \\
P3.2 & Boundary respect (P3.1 fail) & boundary respect by chosen internal entry (LLM, when P3.1 fails)                     & ---                            & LLM semantic    & LLM rejects boundary respect \\
P3.3 & Boundary respect (SAF)    & boundary respect over source (LLM, when call graph unavailable)                         & ---                            & LLM semantic    & LLM rejects boundary respect \\
\midrule
\multicolumn{6}{@{}l}{\textit{\textbf{Principle 4: Entry Point Adequacy} --- entry reaches a security-relevant target}} \\
\midrule
P4.1 & Entry adequacy            & entry reach to core $\land$ $\textit{danger}(e) > 0$ (call graph)                       & ---                            & static call graph & reach $= \emptyset$ or $\textit{danger}(e) = 0$ \\
P4.2 & Entry adequacy (SAF)      & entry adequacy on security-relevant target (LLM, when call graph unavailable)           & ---                            & LLM semantic    & LLM declares entry inadequate \\
\bottomrule
\end{tabular}

\raggedright\footnotesize
\emph{SAF} = static-analysis fallback (e.g., non-C/C++ projects without a precomputed call graph); the LLM takes over the corresponding judgment semantically.

${}^{*}$~For P1.3, either oracle (ASan or LSan) firing on the two-input sequential probe counts as a violation; matches $O_\text{seq}$ in \S\ref{sec:design:principles}.

${}^{\dagger}$~\emph{LLM semantic} = the agent autonomously explores source via \texttt{code\_view} tools (\texttt{read\_file} / \texttt{list\_directory} / \texttt{search\_files}) and judges the property by reasoning; no mechanical verdict from a sanitizer or call-graph computation.
\end{table*}

\input{sections/table_benchmark}

\begin{figure*}[t]
\centering
\includegraphics[width=0.5\textwidth]{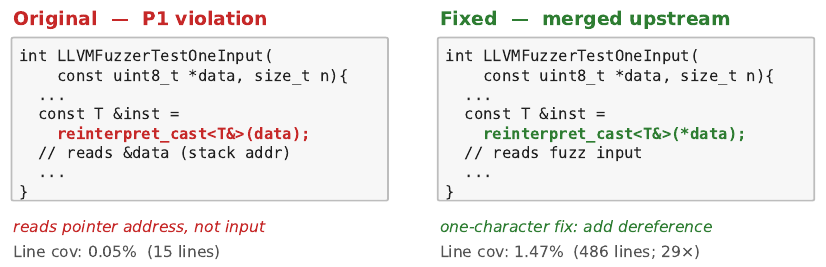}\hspace{-2pt}
\includegraphics[width=0.5\textwidth]{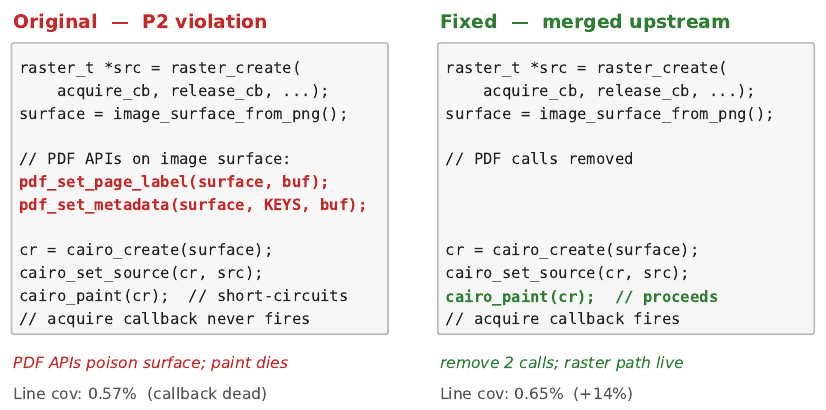}
\caption{Representative P1 and P2 violations identified and repaired by the agent.
\textbf{Top}: harfbuzz \texttt{hb\_set\_fuzzer} (P1)---a \texttt{reinterpret\_cast} on a pointer variable reads its stack address instead of the fuzz input;
one-character fix (\texttt{(data)}$\to$\texttt{(*data)}) restores input flow and raises line coverage $29\times$.
\textbf{Bottom}: cairo \texttt{raster\_fuzzer} (P2)---two PDF-only APIs called on an image surface set \texttt{CAIRO\_STATUS\_SURFACE\_TYPE\_MISMATCH}, causing \texttt{cairo\_paint()} to short-circuit and the acquire/release callbacks to never fire;
removing the two calls makes the raster-source path reachable (merged upstream).}
\label{fig:p1p2_examples}
\end{figure*}

\begin{figure*}[t]
\centering
\includegraphics[width=0.5\textwidth]{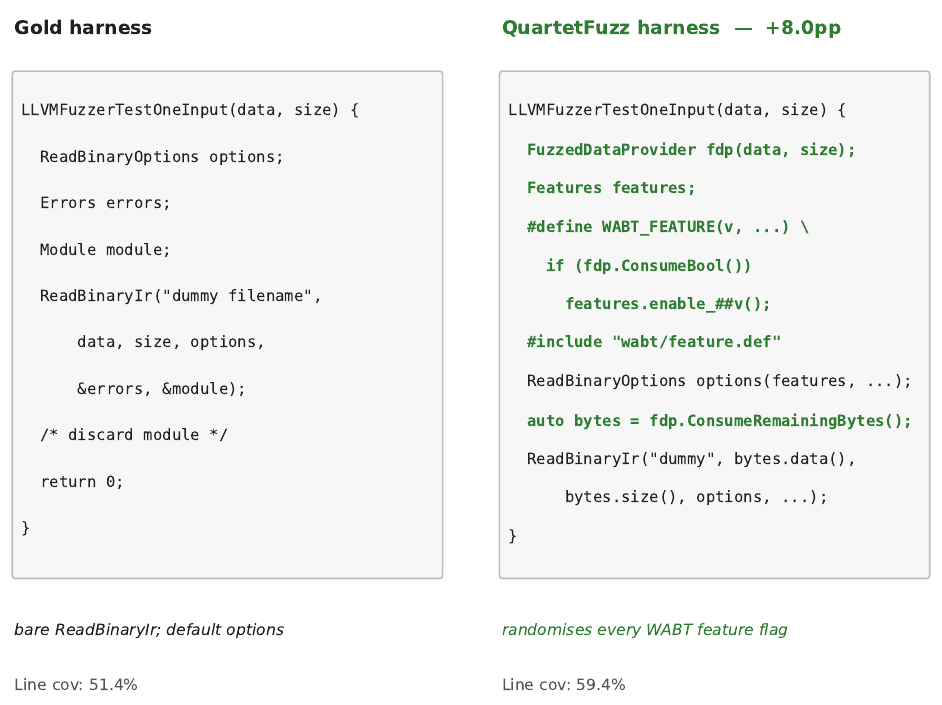}\hspace{-2pt}
\includegraphics[width=0.5\textwidth]{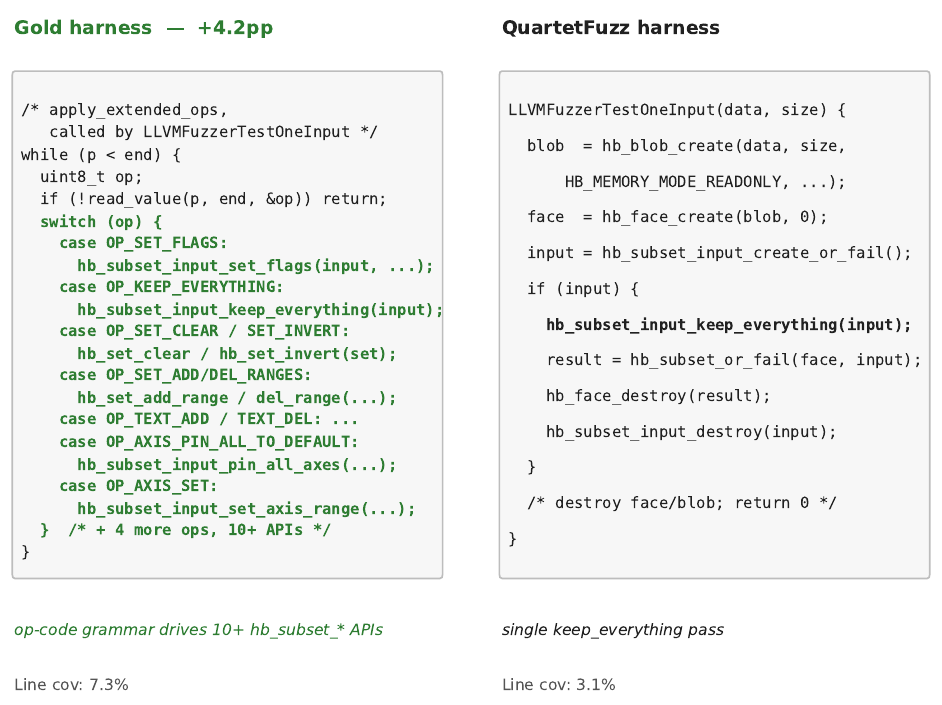}
\caption{Representative RQ3 coverage win and loss vs.\ gold on shared entries.
\textbf{Left}: wabt/wasm2wat ($+8.0$pp)---QF wraps the \texttt{ReadBinaryIr} call with a \texttt{FuzzedDataProvider} that randomises every WABT feature flag, exercising parser paths gold's default-options call never hits.
\textbf{Right}: harfbuzz/subset ($-4.2$pp)---gold parses an op-code grammar that drives 10+ \texttt{hb\_subset\_*} APIs (axis pinning, set clear/invert, range add/del, $\ldots$);
QF runs a single \texttt{keep\_everything} pass, missing the breadth.}
\label{fig:rq3_winloss}
\end{figure*}

\begin{table*}[t]
\caption{RQ1 audit corpus: 70 OSS-Fuzz C/C++ projects, 586 unique fuzzer source files.
\textbf{Stars} = GitHub stars;
\textbf{Fz} = fuzzer source files audited;
\textbf{P1/P2} = principle violations identified;
\textbf{Sub} = issues submitted upstream;
\textbf{Conf/Fix} = confirmed / fixed by maintainers.
Bold project names indicate at least one issue reported upstream.}
\label{tab:rq1_projects}
\scriptsize
\centering
\begin{tabular}{@{}rlrrrrrc|rlrrrrrc@{}}
\toprule
\# & Project & Stars & Fz & P1 & P2 & Sub & Cf/Fx & \# & Project & Stars & Fz & P1 & P2 & Sub & Cf/Fx \\
\midrule
1 & \textbf{apache-httpd} & 3.9k & 7 & 2 & 0 & 2 & 2/2 & 36 & libical & 348 & 3 & 0 & 0 & 0 & 0/0 \\
2 & assimp & 12.9k & 7 & 0 & 0 & 0 & 0/0 & 37 & libjxl & 3.5k & 11 & 0 & 0 & 0 & 0/0 \\
3 & \textbf{binutils} & — & 15 & 1 & 0 & 1 & 1/1 & 38 & libmicrohttpd2 & — & 9 & 0 & 0 & 0 & 0/0 \\
4 & \textbf{boost} & 8.4k & 14 & 2 & 0 & 2 & 1/1 & 39 & \textbf{libpcap} & 3.1k & 5 & 0 & 1 & 1 & 1/1 \\
5 & \textbf{botan} & 3.2k & 34 & 1 & 0 & 1 & 1/1 & 40 & libplist & 616 & 4 & 0 & 0 & 0 & 0/0 \\
6 & brotli & 14.7k & 1 & 0 & 0 & 0 & 0/0 & 41 & \textbf{libpng} & 1.6k & 5 & 1 & 2 & 2 & 2/2 \\
7 & \textbf{bzip2} & 56 & 4 & 2 & 1 & 2 & 2/2 & 42 & libraw & 1.5k & 1 & 0 & 0 & 0 & 0/0 \\
8 & c-ares & 2.1k & 4 & 0 & 0 & 0 & 0/0 & 43 & \textbf{libssh2} & 1.5k & 1 & 1 & 0 & 1 & 1/1 \\
9 & \textbf{cairo} & — & 4 & 0 & 1 & 1 & 1/1 & 44 & libxslt & 71 & 2 & 0 & 0 & 0 & 0/0 \\
10 & curl & 41.3k & 3 & 0 & 0 & 0 & 0/0 & 45 & \textbf{libyaml} & 1.1k & 9 & 1 & 0 & 1 & 1/1 \\
11 & dnsmasq & — & 5 & 0 & 0 & 0 & 0/0 & 46 & llamacpp & 105.0k & 8 & 0 & 0 & 0 & 0/0 \\
12 & draco & 7.2k & 4 & 0 & 0 & 0 & 0/0 & 47 & mbedtls & 6.6k & 8 & 0 & 0 & 0 & 0/0 \\
13 & duckdb & 37.6k & 1 & 0 & 0 & 0 & 0/0 & 48 & \textbf{ndpi} & 4.4k & 61 & 1 & 0 & 1 & 1/1 \\
14 & \textbf{elfutils} & — & 3 & 1 & 1 & 2 & 0/0 & 49 & nettle & — & 7 & 0 & 0 & 0 & 0/0 \\
15 & expat & 1.3k & 2 & 0 & 0 & 0 & 0/0 & 50 & \textbf{njs} & 1.6k & 1 & 1 & 0 & 1 & 1/1 \\
16 & ffmpeg & 59.1k & 6 & 0 & 0 & 0 & 0/0 & 51 & \textbf{opencv} & 87.2k & 9 & 2 & 1 & 3 & 1/1 \\
17 & fftw3 & 3.1k & 1 & 0 & 0 & 0 & 0/0 & 52 & openexr & 1.8k & 2 & 0 & 0 & 0 & 0/0 \\
18 & file & 1.6k & 3 & 0 & 0 & 0 & 0/0 & 53 & \textbf{openssh} & 3.8k & 11 & 1 & 0 & 1 & 0/0 \\
19 & freerdp & 13.1k & 7 & 0 & 0 & 0 & 0/0 & 54 & \textbf{openssl} & 30.0k & 30 & 1 & 1 & 2 & 2/1 \\
20 & \textbf{gdk-pixbuf} & 33 & 5 & 3 & 2 & 3 & 3/3 & 55 & openthread & 3.9k & 6 & 0 & 0 & 0 & 0/0 \\
21 & \textbf{ghostscript} & 218 & 17 & 1 & 0 & 1 & 1/1 & 56 & \textbf{openvpn} & 13.5k & 12 & 2 & 0 & 2 & 2/2 \\
22 & glslang & 3.5k & 1 & 0 & 0 & 0 & 0/0 & 57 & php & 40.0k & 10 & 0 & 0 & 0 & 0/0 \\
23 & \textbf{harfbuzz} & 5.6k & 6 & 1 & 0 & 1 & 1/1 & 58 & pjsip & 2.6k & 17 & 0 & 0 & 0 & 0/0 \\
24 & hwloc & 692 & 1 & 0 & 0 & 0 & 0/0 & 59 & pugixml & 4.5k & 2 & 0 & 0 & 0 & 0/0 \\
25 & \textbf{icu} & 3.5k & 33 & 5 & 4 & 7 & 7/0 & 60 & quickjs & 10.6k & 3 & 0 & 0 & 0 & 0/0 \\
26 & igraph & 2.0k & 26 & 0 & 0 & 0 & 0/0 & 61 & ruby & 23.5k & 10 & 0 & 0 & 0 & 0/0 \\
27 & imagemagick & 16.2k & 9 & 0 & 0 & 0 & 0/0 & 62 & strongswan & 2.8k & 8 & 0 & 0 & 0 & 0/0 \\
28 & iperf & 8.4k & 2 & 0 & 0 & 0 & 0/0 & 63 & \textbf{tidy-html5} & 2.9k & 6 & 1 & 1 & 2 & 2/2 \\
29 & \textbf{jq} & 34.5k & 6 & 1 & 0 & 1 & 1/1 & 64 & \textbf{unbound} & 4.4k & 5 & 1 & 0 & 1 & 1/1 \\
30 & json-c & 3.3k & 4 & 0 & 0 & 0 & 0/0 & 65 & wabt & 8.0k & 5 & 0 & 0 & 0 & 0/0 \\
31 & \textbf{lcms} & 714 & 15 & 5 & 2 & 5 & 4/3 & 66 & \textbf{wamr} & 5.9k & 2 & 0 & 1 & 1 & 1/1 \\
32 & \textbf{libarchive} & 3.5k & 25 & 1 & 0 & 1 & 1/1 & 67 & woff2 & 1.8k & 2 & 0 & 0 & 0 & 0/0 \\
33 & libcoap & 903 & 9 & 0 & 0 & 0 & 0/0 & 68 & yajl-ruby & 1.5k & 1 & 0 & 0 & 0 & 0/0 \\
34 & libgit2 & 10.4k & 8 & 0 & 0 & 0 & 0/0 & 69 & \textbf{zlib} & 6.8k & 12 & 4 & 0 & 4 & 3/2 \\
35 & libheif & 2.2k & 4 & 0 & 0 & 0 & 0/0 & 70 & zopfli & 3.6k & 2 & 0 & 0 & 0 & 0/0 \\
\midrule
\multicolumn{16}{@{}l}{\textbf{Total:} 70 projects (28 in bold with upstream reports), 586 harnesses, 53 P1/P2 reports submitted; 45 confirmed, 35 fixed.} \\
\bottomrule
\end{tabular}

\vspace{4pt}
\raggedright\footnotesize
One additional report (elfutils \texttt{fuzz-dwfl-core}) was withdrawn after a maintainer noted that \texttt{dwfl\_end} and \texttt{elf\_end} already handle NULL gracefully, making the added NULL checks unnecessary---our only P1 false positive, not counted in the 53 above.
\end{table*}

\begin{table*}[t]
\caption{\scriptsize Quality assurance comparison of LLM-based harness generators for C/C++ libraries. Rust-targeting work (deepSURF~\cite{deepsurf}) is discussed in \S\ref{sec:related}.}
\label{tab:llm_comparison}
\tiny
\centering
\begin{tabular}{@{}lcc cccc cc cc c@{}}
\toprule
 & & & \multicolumn{4}{c}{\textbf{Correctness}} & \multicolumn{2}{c}{\textbf{Verification}} & \multicolumn{2}{c}{\textbf{Timing}} & \\
\cmidrule(lr){4-7} \cmidrule(lr){8-9} \cmidrule(lr){10-11}
\textbf{Tool} & \textbf{Venue} & \textbf{Analysis Scope} & \textbf{Compil.} & \textbf{Semantic} & \textbf{API} & \textbf{Entry Pt.} & \textbf{Runtime} & \textbf{Coverage} & \textbf{Pre} & \textbf{Post} & \textbf{Code} \\
\midrule
PromptFuzz~\cite{promptfuzz}      & CCS\,'24   & Headers; AST/CFG          & \ding{51} & \ding{55} & \ding{55} & \ding{55} & \ding{51} & \ding{51} & \ding{55} & \ding{51} & \ding{51} \\
PromeFuzz~\cite{promefuzz}        & CCS\,'25   & AST + consumer call graph + RAG & \ding{51} & \ding{55} & \ding{55} & \ding{55} & \ding{51} & \ding{51} & \ding{55} & \ding{51} & \ding{51} \\
CKGFuzzer~\cite{ckgfuzzer}       & ICSE-C\,'25 & Tree-sitter + CodeQL KG   & \ding{51} & \ding{55} & \ding{55} & \ding{55} & \ding{51} & \ding{51} & \ding{55} & \ding{51} & \ding{51} \\
OSS-Fuzz-Gen~\cite{ossfuzzgen}   & Google     & Headers, Introspector     & \ding{51} & \ding{55} & \ding{55} & \ding{55} & \ding{51} & \ding{51} & \ding{55} & \ding{51} & \ding{51} \\
HarnessAgent~\cite{harnessagent}  & arXiv\,'25 & LSP, Tree-sitter         & \ding{51} & \ding{55} & \ding{55} & \ding{55} & \ding{51} & \ding{51} & \ding{55} & \ding{51} & \ding{55}$^\dagger$ \\
Scheduzz~\cite{scheduzz}          & arXiv\,'25 & Headers; Prolog solver    & \ding{51} & \ding{55} & \ding{51} & \ding{55} & \ding{51} & \ding{51} & \ding{51} & \ding{51} & \ding{55} \\
\midrule
\textbf{QuartetFuzz}               &            & \textbf{Full codebase, call graph} & \textbf{\ding{51}} & \textbf{\ding{51}} & \textbf{\ding{51}} & \textbf{\ding{51}} & \textbf{\ding{51}} & \textbf{\ding{51}} & \textbf{\ding{51}} & \textbf{\ding{51}} & \textbf{\ding{51}} \\
\bottomrule
\end{tabular}

\vspace{4pt}
\raggedright\scriptsize
\textbf{Compil.}: language correctness and build script validity.
\textbf{Semantic}: \emph{pre-fuzz} analysis of the harness source for logic errors, resource management, and harness-introduced bugs (post-fuzz crash root-cause classification is captured by the \textbf{Post} column).
Tools that only check API existence (anti-hallucination) or argument types pre-fuzz, without reading the harness body for logic/resource/lifecycle issues, are marked \ding{55}.
\textbf{API}: explicit runtime call-order/lifecycle/protocol verification (Scheduzz uses Prolog-based constraint solving;
ours uses LLM-driven evidence-grounded protocol reports).
Tools that retrieve API context (signatures, definitions, headers, RAG) without verifying call protocol are marked \ding{55}.
\textbf{Entry Pt.}: security boundary respect and attack-surface adequacy.
\textbf{Pre}: checked before fuzzing. \textbf{Post}: checked after fuzzing.
\textbf{Code}: artifact publicly available for empirical comparison at our submission time.
$^\dagger$~HarnessAgent provides an anonymous artifact at submission via 4open.science but is not packaged for re-execution on our benchmark.
\end{table*}

\input{sections/table_full_multi}

\input{sections/table_ablation_full}

\begin{table*}[t]
\caption{All 44 submitted vulnerability reports across 24 projects: 42 from RQ5 deployment (B--D) and 2 from RQ1 audit-fix (A, latent library bugs surfaced while repairing harness violations).
\textbf{(A)}~RQ1 audit-fix; \textbf{(B)}~C/C++ generation; \textbf{(C)}~Java/JS (LLM-only LG ranking); \textbf{(D)}~rejected by maintainers.
\textit{Danger (Rank)} = per-LG danger score (Eq.~\ref{eq:danger}) and rank among 5 candidates.
\textit{SAST}: \textbf{J}~=~Joern, \textbf{TS}~=~tree-sitter; \textit{SAF} = both backends fail on indirect dispatch (vtable / fn ptr), rank by LLM judgment.
Status: \textsc{fix}ed, \textsc{conf}irmed, \textsc{sub}mitted, \textsc{fp}.}
\label{tab:vulns_projects}
\scriptsize
\centering
\begin{tabular}{@{}rlcrp{0.42\textwidth}p{0.09\textwidth}cc@{}}
\toprule
\# & \textbf{Project} & \textbf{SAST} & \textbf{Danger (Rank)} & \textbf{Bug-finding LG (entry)} & \textbf{Bug Type} & \textbf{CVSS} & \textbf{Status} \\
\midrule
\multicolumn{8}{@{}l}{\textit{(A) Bugs found by fixing existing harness violations (RQ1)}} \\
\midrule
 1 & openssl    & -- & -- & P2 fix: call-order repair in provider fuzzer (\texttt{EVP\_EncryptInit\_ex2}) & stack overread & 7.5 (H) & fix \\
 2 & tidy-html5 & -- & -- & P2 fix: added missing \texttt{tidyCleanAndRepair} call & memory leak & 5.5 (M) & fix \\
\midrule
\multicolumn{8}{@{}l}{\textit{(B) C/C++ projects --- bugs found by directly running generated harnesses}} \\
\midrule
 3 & \textbf{ICU} & J & 121.9 (1) & Transliteration rule compilation (\texttt{Transliterator::createFromRules}) & UAF & 7.5 (H) & fix \\
 4 & opencv      & J &  94.2 (1) & YAML/XML FileStorage parsing (\texttt{cv::FileStorage::open}) & heap overflow & 3.3 (L) & fix \\
 5 & \textbf{openh264}    & J &  20.7 (1) & H.264 bitstream decoding (\texttt{ISVCDecoder::DecodeFrame2}) & heap overflow & 7.5 (H) & conf \\
 6 & \textbf{ghidra}      & J &  14.6 (1) & C++ ABI demangling (\texttt{cplus\_demangle}) & OOM & 5.5 (M) & fix \\
 7 &             & J &  14.6 (1) & Rust v0 demangling (\texttt{rust\_demangle}) & OOM & 5.5 (M) & fix \\
 8 & vorbis      & J &   8.5 (1) & Ogg/Vorbis header parsing (\texttt{vorbis\_synthesis\_headerin}) & memory leak & 5.5 (M) & sub \\
 9 &             & J &   8.5 (1) & Vorbis residue decoding (\texttt{vorbis\_synthesis\_blockin}) & divide-by-zero & 7.5 (H) & sub \\
 10 & capnproto   & J &   7.9 (1) & Cap'n Proto text codec (\texttt{capnp::TextCodec::decode}) & stack overflow & 7.5 (H) & sub \\
 11 & imagemagick$^\star$ & J &  13.2 (2) & MSL script decoding (\texttt{ReadMSLImage}) & NULL deref & 5.5 (M) & fix (CVE-2026-xxxxx) \\
 12 &             $^\star$ & J &  13.2 (2) & MSL script decoding (\texttt{ReadMSLImage}) & stack overflow & 5.5 (M) & fix (CVE-2026-xxxxx) \\
 13 & tidy-html5  & J &   5.8 (1) & Document clean/repair (\texttt{tidyCleanAndRepair}) & memory leak & 5.5 (M) & sub \\
 14 &             & J &   5.8 (2) & Error buffer accumulation (\texttt{tidyErrorSummary}) & OOM & 5.5 (M) & sub \\
 15 &             & J &   5.8 (1) & Generated-doc cleanup (\texttt{TidyGDocClean}) & UAF & 7.5 (H) & sub \\
 16 & \textbf{fwupd}       & J &   4.0 (2) & Logitech RDFU firmware parsing (\texttt{fu\_logitech\_rdfu\_write\_firmware}) & stack overflow & 5.5 (M) & fix \\
 17 & freetype    & J &   1.0 (3) & Glyph bitmap copy (\texttt{FT\_Bitmap\_Copy}) & UAF & 9.1 (C) & sub \\
 18 & libheif     & J &   2.1 (3) & HEIF image crop (\texttt{heif\_image\_crop}) & heap overflow & 7.5 (H) & sub \\
 19 & \textbf{opc-ua}      & J &   3.3 (1) & PubSub JSON configuration decode (\texttt{UA\_PubSub\_decodeJson}) & assertion & 5.5 (M) & conf \\
 20 &             & J &   3.3 (1) & EventFilter parsing (\texttt{UA\_EventFilter\_parse}) & NULL deref & 5.5 (M) & fix \\
 21 & libvpx      & J &  14.2 (1) & Y4M input parsing (\texttt{y4m\_input\_open}) & integer overflow & 7.5 (H) & sub \\
 22 & \textbf{mongoose}    & J &  28.6 (2) & HTTP pattern matching (\texttt{mg\_match}) & heap overflow & 5.5 (M) & fix \\
 23 & \textbf{net-snmp}    & J &   2.8 (4) & VACM config parsing (\texttt{vacm\_parse\_config\_group}) & NULL deref & 5.5 (M) & fix \\
 24 & rapidjson   & J &  18.0 (4) & JSON Schema regex validation (\texttt{SchemaValidator::Validate}) & assertion & 7.5 (H) & sub \\
 25 & libwebp     & J &   5.1 (5) & WebP container assembly (\texttt{WebPMuxAssemble}) & NULL deref & 7.5 (H) & fix \\
 26 &             & J &   2.8 (2) & SharpYuv conversion (\texttt{SharpYuvConvert}, \texttt{FixedPointInterpolation}) & out-of-bounds read & 5.5 (M) & sub \\
 27 & libaom      & TS &   2.3 (3) & AV1 rate control layer-context restore (\texttt{AV1RateControlRTC::ComputeQP}) & heap overflow & 7.5 (H) & conf \\
 28 & libvpx      & TS &   SAF (3) & VP9 complexity-adaptive quantisation (\texttt{vpx\_codec\_encode}) & integer overflow & 7.5 (H) & conf \\
 29 &             & TS &   SAF (3) & VP9 encoder midstream reconfiguration (\texttt{vpx\_codec\_enc\_config\_set}) & heap overflow & 7.5 (H) & conf \\
 30 & \textbf{flatbuffers} & TS &   SAF (4) & FlexBuffers accessor traversal (\texttt{flexbuffers::Reference::ToString}) & heap overflow & 7.5 (H) & conf \\
 31 &             & TS &   5.7 (2) & Reflection verifier field deref (\texttt{flatbuffers::Verify}, \texttt{GetFieldT}) & heap overflow & 5.5 (M) & fix \\
 32 &             & TS &   1.0 (3) & Binary code generation (\texttt{GenerateBinary}) & NULL deref & 5.5 (M) & conf \\
\midrule
\multicolumn{8}{@{}l}{\emph{Bugs below additionally required seed generation:}} \\
\midrule
 33 & \textbf{mongoose}$^\star$    & J &  28.6 (2) & HTTP request parser (\texttt{mg\_match}, \texttt{s.buf[j]} path) & heap overflow & 5.5 (M) & fix (CVE-2026-xxxxx) \\
 34 &             & J &  28.6 (2) & HTTP request parser (\texttt{mg\_match}, write path)            & heap overflow & 5.5 (M) & fix \\
 35 & \textbf{simdutf}     & TS &   SAF (3) & UTF-16\,$\to$\,UTF-8 safe convert (\texttt{convert\_utf16\_to\_utf8\_safe}) & heap overflow & 7.5 (H) & fix \\
 36 & \textbf{fwupd}       & J &   4.0 (2) & JSON manifest loader (\texttt{fwupd\_json\_parser\_load\_array})            & stack overflow & 5.5 (M) & fix \\
 37 & \textbf{flatbuffers} & TS &   7.8 (1) & FBS schema parser (\texttt{FlatBufferBuilder::Finish})                    & NULL deref     & 5.5 (M) & conf \\
\midrule
\multicolumn{8}{@{}l}{\textit{(C) Non-C/C++ projects (Java and JavaScript, LLM-only Logic Group generation \& ranking)}} \\
\midrule
 38 & \textbf{pdfbox}  & -- & SAF (1) & Inline-image decode array (\texttt{PDInlineImage.getDecode})                    & type confusion     & 5.5 (M) & fix \\
 39 &         & -- & SAF (2) & CMap parser (\texttt{CMapParser.increment})                                      & heap overflow      & 5.5 (M) & fix \\
 40 &         & -- & SAF (3) & PFB font parser (\texttt{PfbParser.parsePfb})                                    & integer overflow   & 5.5 (M) & fix \\
 41 & \textbf{graaljs} & -- & SAF (1) & Intl locale validation (\texttt{IntlUtil.validateAndCanonicalizeLanguageTag}) & uncaught exception & 5.5 (M) & fix \\
 42 &         & -- & SAF (2) & RegExp char-class parse (\texttt{RegexLexer.parseCharClassAtomCodePoint})       & uncaught exception & 5.5 (M) & conf \\
\midrule
\multicolumn{8}{@{}l}{\textit{(D) False-positive reports}} \\
\midrule
F1 & \textbf{mongoose} & J & 32.5 (1) & MQTT property iteration (\texttt{mg\_mqtt\_next}) & heap overflow & -- & FP \\
F2 & \textbf{bluez} & TS & 4.4 (3) & MGMT TLV parsing (\texttt{mgmt\_tlv\_list\_load\_from\_buf})           & heap overflow & -- & FP \\
\midrule
\multicolumn{8}{@{}p{0.95\textwidth}}{\textbf{Summary:} 44 total = 42 from RQ5 deployment (20 fixed, 9 confirmed, 11 awaiting, 2 FP) + 2 from RQ1 audit-fix (both fixed).  CVSS v3.1: (C)rit, (H)igh, (M)ed, (L)ow.} \\
\bottomrule
\end{tabular}

\raggedright\footnotesize
\textbf{SAST} = static-analysis backend used to derive the danger score: J~=~Joern (preferred for C/C++), TS~=~tree-sitter fallback (template-heavy C++, Java, JavaScript).
SAF (rows~28--29) = the encoder LG's reach is truncated by C++ vtable indirect dispatch that neither Joern nor tree-sitter resolves; rank tied at danger~$=0$, decided by LLM judgment.
SAF (rows~38--42, Java/JS) = the C/C++ unsafe-keyword set in Eq.~\ref{eq:danger} does not transfer; rank by LLM only.

``--'' = n/a. All CVSS scores are self-estimated by the authors per CVSS v3.1~\cite{cvssv3}.

$^\star$\,Assigned CVE. \textbf{Bold} = direct maintainer collaboration.
\end{table*}

%% file: sections/table_benchmark.tex
\begin{table*}[t]
\caption{Evaluation dataset: 100 gold-standard harnesses from 39 OSS-Fuzz C/C++ projects, all verified P1--P4 clean. \textbf{Target} = the API that consumes the fuzz-byte input. \textbf{Ver} = project commit hash at experiment time (April 2026). \textbf{LOC} = lines of code in the gold harness source.}
\label{tab:benchmark_full}
\scriptsize
\centering
\begin{tabular}{@{}rllllr|rllllr@{}}
\toprule
\# & Project & Fuzzer & Target & Ver & LOC & \# & Project & Fuzzer & Target & Ver & LOC \\
\midrule
1 & apache-httpd & fuzz\_addr\_parse & \texttt{apr\_parse\_addr\_port} & \texttt{1504691} & 37 & 51 & libxslt & xpath & \texttt{xsltFuzzXPath} & \texttt{35323d6} & 21 \\
2 & apache-httpd & fuzz\_tokenize & \texttt{apr\_tokenize\_to\_argv} & \texttt{1504691} & 34 & 52 & libxslt & xslt & \texttt{xsltFuzzXslt} & \texttt{35323d6} & 22 \\
3 & binutils & fuzz\_as & \texttt{perform\_an\_assembl..} & \texttt{9b9cbb0} & 77 & 53 & libyaml & libyaml\_loader\_fuz.. & \texttt{yaml\_parser\_load} & \texttt{840b65c} & 51 \\
4 & binutils & fuzz\_disassemble & \texttt{disassembler} & \texttt{9b9cbb0} & 110 & 54 & libyaml & libyaml\_scanner\_fu.. & \texttt{yaml\_parser\_scan} & \texttt{840b65c} & 51 \\
5 & boost & boost\_graph\_graphv.. & \texttt{read\_graphviz} & \texttt{1a80576} & 57 & 55 & llamacpp & fuzz\_json\_to\_grammar & \texttt{json\_schema\_to\_gra..} & \texttt{cf8b0db} & 29 \\
6 & boost & inforead\_fuzzer & \texttt{pt::read\_info} & \texttt{1a80576} & 46 & 56 & mbedtls & fuzz\_pkcs7 & \texttt{mbedtls\_pkcs7\_pars..} & \texttt{391af7c} & 21 \\
7 & boost & iniread\_fuzzer & \texttt{pt::read\_ini} & \texttt{1a80576} & 46 & 57 & mbedtls & fuzz\_x509crl & \texttt{mbedtls\_x509\_crl\_p..} & \texttt{391af7c} & 40 \\
8 & boost & jsonread\_fuzzer & \texttt{pt::read\_json} & \texttt{1a80576} & 46 & 58 & mbedtls & fuzz\_x509crt & \texttt{mbedtls\_x509\_crt\_p..} & \texttt{391af7c} & 40 \\
9 & boost & xmlread\_fuzzer & \texttt{pt::read\_xml} & \texttt{1a80576} & 64 & 59 & mbedtls & fuzz\_x509csr & \texttt{mbedtls\_x509\_csr\_p..} & \texttt{391af7c} & 40 \\
10 & brotli & decode\_fuzzer & \texttt{BrotliDecoderDecom..} & \texttt{ab685df} & 63 & 60 & ndpi & fuzz\_community\_id & \texttt{ndpi\_flowv4\_flow\_h..} & \texttt{315a705} & 58 \\
11 & curl & curl\_fuzzer\_ftp & \texttt{fuzz\_handle\_transfer} & \texttt{70a1595} & 584 & 61 & ndpi & fuzz\_dga & \texttt{ndpi\_check\_dga\_name} & \texttt{315a705} & 46 \\
12 & curl & fuzz\_url & \texttt{curl\_url\_set} & \texttt{70a1595} & 55 & 62 & ndpi & fuzz\_ds\_tree & \texttt{ndpi\_tsearch} & \texttt{315a705} & 99 \\
13 & draco & mesh\_decoder\_fuzzer & \texttt{draco::Decoder::De..} & \texttt{77e616e} & 29 & 63 & ndpi & categories & \texttt{load\_categories\_fi..} & \texttt{315a705} & 24 \\
14 & draco & mesh\_decoder\_witho.. & \texttt{draco::Decoder::De..} & \texttt{77e616e} & 30 & 64 & ndpi & category & \texttt{load\_category\_file..} & \texttt{315a705} & 24 \\
15 & draco & pc\_decoder\_fuzzer & \texttt{draco::Decoder::De..} & \texttt{77e616e} & 29 & 65 & ndpi & fuzz\_filecfg\_config & \texttt{load\_config\_file\_fd} & \texttt{315a705} & 24 \\
16 & draco & pc\_decoder\_without.. & \texttt{draco::Decoder::De..} & \texttt{77e616e} & 30 & 66 & ndpi & malicious\_ja4 & \texttt{load\_malicious\_ja4..} & \texttt{315a705} & 24 \\
17 & fftw3 & fftw3\_fuzzer & \texttt{fftw\_plan\_dft\_1d} & \texttt{6caf8ce} & 40 & 67 & ndpi & malicious\_sha1 & \texttt{load\_malicious\_sha..} & \texttt{315a705} & 42 \\
18 & freerdp & TestFuzzCodecs & \texttt{xcrush\_decompress} & \texttt{0c31662} & 467 & 68 & ndpi & protocols & \texttt{load\_protocols\_fil..} & \texttt{315a705} & 24 \\
19 & glslang & compile\_fuzzer & \texttt{glslang::TShader s..} & \texttt{aa8e19e} & 32 & 69 & ndpi & risk\_domains & \texttt{load\_risk\_domain\_f..} & \texttt{315a705} & 24 \\
20 & harfbuzz & hb-shape-fuzzer & \texttt{hb\_shape} & \texttt{e8fbf40} & 375 & 70 & ndpi & fuzz\_process\_packet & \texttt{ndpi\_detection\_pro..} & \texttt{315a705} & 41 \\
21 & harfbuzz & hb-subset-fuzzer & \texttt{hb\_subset\_or\_fail} & \texttt{e8fbf40} & 356 & 71 & opencv & imread\_fuzzer & \texttt{cv::imread} & \texttt{9f101a1} & 16 \\
22 & hwloc & hwloc\_fuzzer & \texttt{hwloc\_encode\_to\_ba..} & \texttt{cfe0433} & 36 & 72 & openexr & openexr\_exrcheck\_f.. & \texttt{checkOpenEXRFile} & \texttt{53cfa83} & 16 \\
23 & icu & calendar\_fuzzer & \texttt{icu::Calendar::cre..} & \texttt{0d84e02} & 136 & 73 & openexr & openexr\_exrcoreche.. & \texttt{checkOpenEXRFile} & \texttt{53cfa83} & 15 \\
24 & icu & date\_time\_pattern\_.. & \texttt{icu::DateTimePatte..} & \texttt{0d84e02} & 58 & 74 & openssh & authopt\_fuzz & \texttt{sshauthopt\_parse} & \texttt{45b30e0} & 33 \\
25 & icu & normalizer2\_fuzzer & \texttt{icu::Normalizer2::..} & \texttt{0d84e02} & 83 & 75 & openssh & kex\_fuzz & \texttt{ssh\_init} & \texttt{45b30e0} & 454 \\
26 & icu & number\_format\_fuzzer & \texttt{icu::NumberFormat:..} & \texttt{0d84e02} & 93 & 76 & openssh & privkey\_fuzz & \texttt{sshkey\_private\_des..} & \texttt{45b30e0} & 20 \\
27 & icu & unicode\_string\_cod.. & \texttt{icu::UnicodeString} & \texttt{0d84e02} & 83 & 77 & openssh & sig\_fuzz & \texttt{sshkey\_verify} & \texttt{45b30e0} & 62 \\
28 & imagemagick & encoder\_mvg\_fuzzer & \texttt{Magick::Image::read} & \texttt{5e318be} & 121 & 78 & openssh & sshsigopt\_fuzz & \texttt{sshsigopt\_parse} & \texttt{45b30e0} & 29 \\
29 & imagemagick & ping\_fuzzer & \texttt{Magick::Image::ping} & \texttt{5e318be} & 50 & 79 & openssl & acert & \texttt{d2i\_X509\_ACERT} & \texttt{087bddc} & 48 \\
30 & iperf & cjson\_fuzzer & \texttt{cJSON\_Parse} & \texttt{896cc42} & 38 & 80 & openssl & asn1parse & \texttt{ASN1\_parse\_dump} & \texttt{087bddc} & 45 \\
31 & jq & jq\_fuzz\_parse & \texttt{jv\_parse} & \texttt{fb59f14} & 21 & 81 & openssl & cms & \texttt{i2d\_CMS\_bio} & \texttt{087bddc} & 55 \\
32 & jq & jq\_fuzz\_parse\_exte.. & \texttt{jv\_parse\_custom\_fl..} & \texttt{fb59f14} & 36 & 82 & openssl & v3name & \texttt{GENERAL\_NAME\_cmp} & \texttt{087bddc} & 45 \\
33 & libcoap & get\_asn1\_tag\_target & \texttt{get\_asn1\_tag} & \texttt{86b7781} & 17 & 83 & php & fuzzer-json & \texttt{php\_json\_yyparse} & \texttt{e07d066} & 61 \\
34 & libcoap & oscore\_conf\_parse\_.. & \texttt{coap\_new\_oscore\_conf} & \texttt{86b7781} & 46 & 84 & php & fuzzer-unserialize & \texttt{php\_var\_unserialize} & \texttt{e07d066} & 68 \\
35 & libcoap & split\_uri\_target & \texttt{coap\_split\_uri} & \texttt{86b7781} & 8 & 85 & php & fuzzer-unserialize.. & \texttt{php\_var\_unserialize} & \texttt{e07d066} & 82 \\
36 & libgit2 & objects\_fuzzer & \texttt{git\_object\_\_from\_raw} & \texttt{1f34e2a} & 49 & 86 & pjsip & fuzz-crypto & \texttt{pj\_base64\_encode} & \texttt{2965cff} & 181 \\
37 & libgit2 & patch\_parse\_fuzzer & \texttt{git\_patch\_from\_buf..} & \texttt{1f34e2a} & 40 & 87 & pjsip & fuzz-dns & \texttt{pj\_dns\_parse\_packet} & \texttt{2965cff} & 78 \\
38 & libical & libical\_fuzzer & \texttt{icalparser\_parse\_s..} & \texttt{276b8bd} & 44 & 88 & pjsip & fuzz-sip & \texttt{pjsip\_parse\_msg} & \texttt{2965cff} & 489 \\
39 & libical & libicalvcard\_fuzzer & \texttt{vcardparser\_parse\_..} & \texttt{276b8bd} & 45 & 89 & pjsip & fuzz-srtp & \texttt{pjmedia\_transport\_..} & \texttt{2965cff} & 201 \\
40 & libjxl & cjxl\_fuzzer & \texttt{EncodeJpegXl} & \texttt{6553831} & 260 & 90 & pugixml & fuzz\_parse & \texttt{pugi::xml\_document..} & \texttt{e56134e} & 14 \\
41 & libjxl & color\_encoding\_fuz.. & \texttt{jxl::ParseDescript..} & \texttt{6553831} & 36 & 91 & pugixml & fuzz\_xpath & \texttt{evaluate\_node\_set} & \texttt{e56134e} & 46 \\
42 & libjxl & fields\_fuzzer & \texttt{jxl::Bundle::Read} & \texttt{6553831} & 107 & 92 & quickjs & fuzz\_compile & \texttt{JS\_Eval} & \texttt{d7ae12a} & 93 \\
43 & libjxl & icc\_codec\_fuzzer & \texttt{jxl::UnpredictICC} & \texttt{6553831} & 115 & 93 & quickjs & fuzz\_eval & \texttt{JS\_Eval} & \texttt{d7ae12a} & 49 \\
44 & libjxl & set\_from\_bytes\_fuz.. & \texttt{jxl::SetFromBytes} & \texttt{6553831} & 77 & 94 & strongswan & fuzz\_crls & \texttt{lib->creds->create} & \texttt{60f4c86} & 41 \\
45 & libpcap & fuzz\_both & \texttt{pcap\_compile} & \texttt{44aa24f} & 113 & 95 & strongswan & fuzz\_ids & \texttt{identification\_cre..} & \texttt{60f4c86} & 34 \\
46 & libpcap & fuzz\_filter & \texttt{pcap\_compile} & \texttt{44aa24f} & 44 & 96 & wabt & wasm2wat\_fuzzer & \texttt{ReadBinaryIr} & \texttt{77a95c8} & 26 \\
47 & libplist & bplist\_fuzzer & \texttt{plist\_from\_bin} & \texttt{dddb76d} & 32 & 97 & yajl-ruby & json\_fuzzer & \texttt{yajl\_parse} & \texttt{6501652} & 104 \\
48 & libplist & jplist\_fuzzer & \texttt{plist\_from\_json} & \texttt{dddb76d} & 32 & 98 & zlib & compress\_fuzzer & \texttt{compress2} & \texttt{f9dd600} & 99 \\
49 & libplist & oplist\_fuzzer & \texttt{plist\_from\_openstep} & \texttt{dddb76d} & 32 & 99 & zlib & zlib\_uncompress2\_f.. & \texttt{uncompress2} & \texttt{f9dd600} & 16 \\
50 & libplist & xplist\_fuzzer & \texttt{plist\_from\_xml} & \texttt{dddb76d} & 32 & 100 & zopfli & zopfli\_deflate\_fuz.. & \texttt{ZopfliDeflate} & \texttt{ccf9f05} & 45 \\
\bottomrule
\end{tabular}
\end{table*}

%% file: sections/table_full_multi.tex
\begin{table*}[t]
\caption{Full coverage comparison (line, branch \%) vs.\ three baselines. G = Gold, O = OFG, P = PromeFuzz, Q = QuartetFuzz; AP = QuartetFuzz Adversarial Probing calls per case ($0$ if the case never produced a Stage-3 binary; Average is over the 96 productive cases). Each harness fuzzed under LibFuzzer for $10\times600$\,s (empty corpus, ASan); per-cell value is the median across the 10 runs of \texttt{llvm-cov} line/branch.}
\label{tab:full_comparison_multi}
\tiny
\centering
\resizebox{\textwidth}{!}{%
\begin{tabular}{@{}rl rrrr|rrrr|r || rl rrrr|rrrr|r @{}}
\toprule
 &  & \multicolumn{4}{c}{Line} & \multicolumn{4}{c}{Branch} & & &  & \multicolumn{4}{c}{Line} & \multicolumn{4}{c}{Branch} & \\
\cmidrule(lr){3-6} \cmidrule(lr){7-10} \cmidrule(lr){14-17} \cmidrule(lr){18-21}
\# & Case & G & O & P & Q & G & O & P & Q & AP & \# & Case & G & O & P & Q & G & O & P & Q & AP \\
\midrule
1 & ap-h/addr\_parse & 1.0 & 1.6 & 98 & 1.6 & 0.5 & 0.9 & 57 & 0.9 & 7 & 51 & libxslt/xpath & 16 & 1.3 & 1.0 & 16 & 14 & 0.4 & 0.5 & 14 & 2 \\
2 & ap-h/tokenize & 1.1 & 1.4 & 98 & 1.4 & 0.5 & 0.9 & 57 & 0.9 & 6 & 52 & libxslt/xslt & 9.8 & 7.8 & 6.5 & 6.8 & 7.8 & 6.2 & 3.6 & 5.2 & 1 \\
3 & binutils/as & 2.3 & 0 & 0.5 & 0 & 1.9 & 0 & 0.5 & 0 & 0 & 53 & libyaml/loader & 78 & 74 & 21 & 78 & 66 & 60 & 10 & 67 & 3 \\
4 & binutils/disasm & 18 & 17 & 0 & 8.6 & 16 & 15 & 0 & 7.0 & 1 & 54 & libyaml/scanner & 71 & 70 & 69 & 71 & 63 & 62 & 65 & 64 & 3 \\
5 & boost/graphviz & 37 & 34 & 0 & 37 & 33 & 32 & 0 & 33 & 2 & 55 & llama/js-grm & 3.4 & 3.3 & 0 & 3.8 & 3.6 & 3.4 & 0 & 3.8 & 3 \\
6 & boost/inforead & 56 & 0 & 0 & 55 & 79 & 0 & 0 & 79 & 2 & 56 & mbedtls/pkcs7 & 1.0 & 0.8 & 1.9 & 3.1 & 0.6 & 0.6 & 1.5 & 2.4 & 1 \\
7 & boost/iniread & 54 & 0 & 0 & 54 & 60 & 0 & 0 & 60 & 4 & 57 & mbedtls/x509crl & 1.5 & 1.5 & 1.6 & 3.7 & 1.3 & 1.3 & 1.5 & 3.0 & 8 \\
8 & boost/jsonread & 62 & 0 & 0 & 62 & 85 & 0 & 0 & 85 & 2 & 58 & mbedtls/x509crt & 1.7 & 1.7 & 1.9 & 3.8 & 1.5 & 1.5 & 1.8 & 3.0 & 5 \\
9 & boost/xmlread & 60 & 0 & 0 & 59 & 68 & 0 & 0 & 64 & 2 & 59 & mbedtls/x509csr & 1.4 & 1.4 & 1.6 & 3.6 & 1.2 & 1.2 & 1.6 & 2.9 & 2 \\
10 & brotli/decode & 77 & 77 & 23 & 80 & 72 & 69 & 17 & 75 & 3 & 60 & ndpi/cm-id & 1.2 & 0.5 & 4.8 & 0.9 & 0.5 & 0.3 & 1.6 & 0.5 & 7 \\
11 & curl/curl-ftp & 5.0 & 0 & 2.7 & 0 & 3.9 & 0 & 0.9 & 0 & 0 & 61 & ndpi/dga & 16 & 16 & 9.5 & 16 & 7.3 & 7.2 & 3.2 & 7.3 & 6 \\
12 & curl/url & 4.0 & 4.0 & 0.5 & 0.4 & 3.2 & 3.1 & 0.3 & 0.4 & 7 & 62 & ndpi/ds\_tree & 1.0 & 0.5 & 0 & 0.6 & 0.3 & 0.3 & 0 & 0.4 & 11 \\
13 & draco/mesh & 11 & 0 & 12 & 20 & 9.7 & 0 & 13 & 19 & 2 & 63 & ndpi/fc-cat & 6.3 & 6.0 & 4.5 & 6.3 & 2.4 & 2.0 & 1.5 & 2.5 & 9 \\
14 & draco/mesh\_nodq & 11 & 0 & 9.8 & 7.3 & 10 & 0 & 11 & 6.3 & 3 & 64 & ndpi/fc-cty & 6.3 & 6.0 & 4.5 & 6.3 & 2.7 & 2.2 & 1.5 & 2.7 & 7 \\
15 & draco/pc & 3.6 & 0 & 11 & 3.4 & 2.5 & 0 & 13 & 3.0 & 2 & 65 & ndpi/fc-cfg & 6.0 & 5.7 & 4.5 & 6.0 & 1.7 & 1.3 & 1.5 & 1.7 & 6 \\
16 & draco/pc\_nodq & 4.2 & 0 & 10 & 3.0 & 3.7 & 0 & 12 & 2.1 & 2 & 66 & ndpi/fc-mj & 5.8 & 5.5 & 4.5 & 5.8 & 1.9 & 1.5 & 1.5 & 1.9 & 9 \\
17 & fftw3/fftw3 & 14 & 21 & 84 & 21 & 31 & 32 & 58 & 32 & 3 & 67 & ndpi/fc-ms & 1.1 & 5.5 & 4.5 & 5.9 & 0.9 & 1.5 & 1.5 & 2.0 & 9 \\
18 & freerdp/Codecs & 14 & 13 & 0 & 14 & 10 & 9.6 & 0 & 10 & 2 & 68 & ndpi/fc-pro & 7.6 & 7.5 & 4.5 & 7.9 & 3.5 & 3.2 & 1.5 & 3.8 & 8 \\
19 & glslang/compile & 23 & 0 & 0.3 & 21 & 21 & 0 & 0.1 & 19 & 3 & 69 & ndpi/fc-rsk & 6.7 & 6.4 & 4.5 & 6.7 & 2.3 & 1.8 & 1.5 & 2.3 & 9 \\
20 & harfbuzz/shape-fuzzer & 19 & 19 & 1.0 & 19 & 23 & 23 & 0.5 & 23 & 2 & 70 & ndpi/proc-pkt & 48 & 47 & 4.8 & 47 & 39 & 38 & 1.5 & 40 & 7 \\
21 & harfbuzz/subset & 7.3 & 8.3 & 1.5 & 3.1 & 4.6 & 5.6 & 0.7 & 2.0 & 3 & 71 & opencv/imread & 4.4 & 0 & 0 & 4.5 & 3.2 & 0 & 0 & 3.4 & 2 \\
22 & hwloc/hwloc & 15 & 6.2 & 7.4 & 14 & 86 & 17 & 73 & 63 & 3 & 72 & openexr/exrcheck & 5.3 & 3.5 & 6.5 & 5.3 & 4.3 & 2.8 & 25 & 4.3 & 2 \\
23 & icu/calendar & 16 & 0 & 4.2 & 0 & 12 & 0 & 2.3 & 0 & 0 & 73 & openexr/corecheck & 5.1 & 3.9 & 6.5 & 5.0 & 4.3 & 3.3 & 25 & 4.2 & 3 \\
24 & icu/dt & 14 & 0 & 4.2 & 14 & 11 & 0 & 2.3 & 11 & 3 & 74 & openssh/authopt & 4.2 & 3.6 & 3.6 & 7.9 & 4.4 & 3.5 & 3.5 & 7.7 & 1 \\
25 & icu/normalizer2 & 13 & 0 & 4.2 & 16 & 11 & 0 & 2.3 & 16 & 5 & 75 & openssh/kex & 9.5 & 20 & 0 & 9.1 & 10 & 20 & 0 & 8.9 & 2 \\
26 & icu/num-fmt & 14 & 0 & 4.2 & 14 & 11 & 0 & 2.3 & 11 & 3 & 76 & openssh/privkey & 6.1 & 4.9 & 5.7 & 6.3 & 4.8 & 4.1 & 4.6 & 4.9 & 2 \\
27 & icu/ustr-cp & 9.0 & 0 & 4.2 & 13 & 6.6 & 0 & 2.3 & 10 & 6 & 77 & openssh/sig & 5.4 & 2.9 & 2.9 & 5.4 & 4.3 & 2.3 & 2.3 & 4.3 & 3 \\
28 & imgmk/encoder\_mvg & 1.0 & 0 & 1.8 & 0.8 & 0.6 & 0 & 1.4 & 0.6 & 2 & 78 & openssh/sshsigopt & 1.3 & 0 & 0 & 1.3 & 1.4 & 0 & 0 & 1.4 & 3 \\
29 & imgmk/ping & 4.9 & 0 & 1.8 & 5.1 & 3.7 & 0 & 1.4 & 4.0 & 4 & 79 & openssl/acert & 2.7 & 2.1 & 3.0 & 2.6 & 2.1 & 1.5 & 2.3 & 1.9 & 2 \\
30 & iperf/cjson & 24 & 24 & 27 & 24 & 27 & 26 & 29 & 26 & 7 & 80 & openssl/asn1parse & 2.2 & 2.2 & 0 & 2.2 & 1.5 & 1.6 & 0 & 1.6 & 3 \\
31 & jq/parse & 3.5 & 3.1 & 3.3 & 3.4 & 3.5 & 3.2 & 3.1 & 3.5 & 2 & 81 & openssl/cms & 2.4 & 2.7 & 0 & 2.4 & 1.8 & 2.1 & 0 & 1.8 & 2 \\
32 & jq/prs-ext & 7.2 & 4.2 & 0 & 6.4 & 7.1 & 4.6 & 0 & 6.4 & 3 & 82 & openssl/v3name & 2.6 & 1.8 & 0 & 2.8 & 2.1 & 1.3 & 0 & 2.0 & 2 \\
33 & libcoap/asn1 & 1.1 & 0 & 0.3 & 0.5 & 0.3 & 0 & 0.1 & 0.3 & 4 & 83 & php/json & 5.0 & 0 & 0 & 5.0 & 1.2 & 0 & 0 & 1.3 & 3 \\
34 & libcoap/oscore & 1.8 & 0 & 0.3 & 1.8 & 1.1 & 0 & 0.1 & 1.1 & 4 & 84 & php/unseria & 4.8 & 0 & 0 & 4.9 & 1.1 & 0 & 0 & 1.2 & 3 \\
35 & libcoap/uri & 1.2 & 0 & 0.3 & 1.8 & 0.8 & 0 & 0.1 & 1.4 & 4 & 85 & php/unsr-h & 4.9 & 0 & 0 & 5.0 & 1.2 & 0 & 0 & 1.2 & 4 \\
36 & libgit2/objects & 2.3 & 1.5 & 10 & 2.3 & 1.8 & 1.1 & 8.1 & 1.8 & 10 & 86 & pjsip/crypto & 54 & 46 & 49 & 57 & 60 & 50 & 72 & 64 & 7 \\
37 & libgit2/patch & 1.5 & 1.2 & 0 & 1.5 & 1.0 & 0.9 & 0 & 1.0 & 10 & 87 & pjsip/dns & 20 & 13 & 38 & 20 & 9.5 & 6.2 & 80 & 9.9 & 2 \\
38 & libical/libical & 12 & 11 & 15 & 13 & 11 & 9.8 & 15 & 12 & 2 & 88 & pjsip/sip & 14 & 1.3 & 49 & 12 & 7.9 & 1.0 & 72 & 6.5 & 1 \\
39 & libical/vcard & 8.5 & 14 & 16 & 0 & 9.1 & 13 & 16 & 0 & 0 & 89 & pjsip/srtp & 22 & 14 & 49 & 22 & 11 & 7.2 & 72 & 11 & 2 \\
40 & libjxl/cjxl & 21 & 0 & 25 & 21 & 15 & 0 & 32 & 15 & 2 & 90 & pugixml/parse & 13 & 0 & 15 & 15 & 13 & 0 & 14 & 16 & 2 \\
41 & libjxl/col-enc & 81 & 0 & 23 & 80 & 97 & 0 & 30 & 96 & 2 & 91 & pugixml/xpath & 42 & 42 & 22 & 43 & 36 & 35 & 22 & 37 & 5 \\
42 & libjxl/fields & 25 & 0 & 0 & 25 & 45 & 0 & 0 & 45 & 1 & 92 & quickjs/compile & 26 & 26 & 18 & 28 & 23 & 22 & 56 & 24 & 3 \\
43 & libjxl/icc & 7.2 & 0 & 25 & 7.2 & 17 & 0 & 32 & 18 & 3 & 93 & quickjs/eval & 25 & 25 & 18 & 27 & 22 & 21 & 56 & 24 & 3 \\
44 & libjxl/frm-byt & 15 & 0 & 23 & 17 & 15 & 0 & 28 & 17 & 2 & 94 & sswan/crls & 11 & 0 & 0 & 11 & 7.0 & 0 & 0 & 6.9 & 2 \\
45 & libpcap/both & 20 & 18 & 0.8 & 19 & 22 & 20 & 0.4 & 21 & 3 & 95 & sswan/ids & 8.0 & 5.9 & 0 & 7.9 & 4.2 & 3.1 & 0 & 5.3 & 3 \\
46 & libpcap/filter & 24 & 25 & 0.9 & 24 & 27 & 26 & 0.5 & 27 & 2 & 96 & wabt/wasm2wat & 51 & 0 & 0 & 59 & 59 & 0 & 0 & 68 & 2 \\
47 & libplist/bplist & 9.9 & 6.4 & 24 & 9.7 & 8.3 & 5.3 & 21 & 8.1 & 2 & 97 & yajl-ruby/json & 69 & 69 & 2.7 & 78 & 53 & 52 & 0.6 & 69 & 2 \\
48 & libplist/jplist & 12 & 6.6 & 22 & 12 & 10 & 4.8 & 19 & 11 & 3 & 98 & zlib/compress & 55 & 55 & 52 & 51 & 45 & 45 & 42 & 42 & 12 \\
49 & libplist/oplist & 10 & 1.3 & 22 & 10 & 8.0 & 0.7 & 19 & 8.0 & 2 & 99 & zlib/uncomp2 & 53 & 53 & 53 & 54 & 50 & 43 & 43 & 49 & 7 \\
50 & libplist/xplist & 11 & 11 & 7.2 & 11 & 8.4 & 9.1 & 5.0 & 8.4 & 2 & 100 & zopfli/deflate & 86 & 83 & 72 & 86 & 70 & 68 & 54 & 70 & 2 \\
\midrule
& \textbf{Average} & \textbf{17.7} & \textbf{10.8} & \textbf{12.5} & \textbf{17.7} & \textbf{17.6} & \textbf{9.2} & \textbf{13.4} & \textbf{17.5} & \textbf{3.77} & \multicolumn{11}{c}{} \\
& \textbf{Productive rate} & \multicolumn{1}{c}{100} & \multicolumn{1}{c}{64} & \multicolumn{1}{c}{74} & \multicolumn{1}{c}{96} & & & & & & \multicolumn{11}{c}{} \\
& \textbf{Avg.\ cost} & \multicolumn{1}{c}{{-}} & \multicolumn{1}{c}{\$2.43} & \multicolumn{1}{c}{\$1.98} & \multicolumn{1}{c}{\$1.65} & & & & & & \multicolumn{11}{c}{} \\
& \textbf{Model} & \multicolumn{4}{c}{claude-sonnet-4-6} & & & & & & \multicolumn{11}{c}{} \\
\cmidrule{1-11}
\end{tabular}}%
\end{table*}

%% file: sections/table_ablation_full.tex
\begin{table*}[t]
\caption{RQ4 ablation: per-case coverage (\%) under four configurations. Each harness fuzzed under LibFuzzer for $10\times600$\,s (empty corpus, ASan); per-cell value is the median across the 10 runs (\texttt{llvm-cov} line/branch). AP = without adversarial probe gate; P2 = without API protocol report; St = without call-graph tools; Gm = full pipeline with Gemini~3.1~Flash. All four are ablations of the full pipeline whose aggregate appears in Table~\ref{tab:rq3_headline}.}
\label{tab:ablation_full}
\tiny
\centering
\resizebox{\textwidth}{!}{%
\begin{tabular}{@{}rl rrrr|rrrr || rl rrrr|rrrr @{}}
\toprule
 & & \multicolumn{4}{c}{Line} & \multicolumn{4}{c}{Branch} & & & \multicolumn{4}{c}{Line} & \multicolumn{4}{c}{Branch} \\
\cmidrule(lr){3-6} \cmidrule(lr){7-10} \cmidrule(lr){13-16} \cmidrule(lr){17-20}
\# & Case & AP & P2 & St & Gm & AP & P2 & St & Gm & \# & Case & AP & P2 & St & Gm & AP & P2 & St & Gm \\
\midrule
1 & ap-h/addr\_par & 1.6 & 1.6 & 1.6 & 1.6 & 0.9 & 0.9 & 0.9 & 0.8 & 51 & libxslt/xpath & 16 & 16 & 15 & 16 & 14 & 13 & 13 & 13 \\
2 & ap-h/tokenize & 1.4 & 1.4 & 1.4 & 1.4 & 0.9 & 0.9 & 0.9 & 0.9 & 52 & libxslt/xslt & 8.6 & 0 & 9.8 & 9.4 & 7.0 & 0 & 8.0 & 7.7 \\
3 & binutils/as & 0 & 0 & 2.4 & 0 & 0 & 0 & 2.0 & 0 & 53 & libyaml/libyaml\_ & 78 & 78 & 78 & 78 & 67 & 66 & 66 & 66 \\
4 & binutils/disassem & 8.3 & 8.9 & 8.9 & 8.4 & 6.7 & 7.2 & 7.2 & 6.8 & 54 & libyaml/libyaml\_ & 71 & 71 & 71 & 71 & 63 & 64 & 63 & 63 \\
5 & boost/graph\_gr & 37 & 37 & 37 & 37 & 33 & 33 & 33 & 33 & 55 & llama/s/json\_t & 3.8 & 3.7 & 3.7 & 3.5 & 3.9 & 3.7 & 3.8 & 3.6 \\
6 & boost/ptree\_in & 0 & 57 & 57 & 55 & 0 & 81 & 81 & 79 & 56 & mbedtls/pkcs7 & 0 & 0.8 & 1.0 & 0 & 0 & 0.5 & 0.8 & 0 \\
7 & boost/ptree\_in & 54 & 54 & 54 & 54 & 60 & 60 & 60 & 60 & 57 & mbedtls/x509crl & 0 & 0 & 1.8 & 0 & 0 & 0 & 1.5 & 0 \\
8 & boost/ptree\_js & 62 & 62 & 62 & 62 & 85 & 86 & 85 & 85 & 58 & mbedtls/x509crt & 0 & 0 & 1.7 & 0 & 0 & 0 & 1.5 & 0 \\
9 & boost/ptree\_xm & 59 & 0 & 59 & 62 & 64 & 0 & 66 & 73 & 59 & mbedtls/x509csr & 0 & 0 & 1.4 & 3.7 & 0 & 0 & 1.2 & 3.0 \\
10 & brotli/decode\_ & 78 & 77 & 77 & 78 & 73 & 72 & 73 & 71 & 60 & ndpi/communit & 0.9 & 0.8 & 0.8 & 0.8 & 0.5 & 0.5 & 0.5 & 0.5 \\
11 & curl/curl\_\_ft & 0 & 0 & 5.0 & 3.6 & 0 & 0 & 3.9 & 2.4 & 61 & ndpi/dga & 16 & 16 & 16 & 16 & 7.2 & 7.2 & 7.2 & 7.2 \\
12 & curl/url & 0.4 & 0.4 & 0.4 & 0.4 & 0.4 & 0.4 & 0.4 & 0.4 & 62 & ndpi/ds\_tree & 0.6 & 0.6 & 0.6 & 0.6 & 0.4 & 0.4 & 0.4 & 0.5 \\
13 & draco/draco\_me & 4.8 & 0 & 6.7 & 6.9 & 4.3 & 0 & 5.5 & 6.0 & 63 & ndpi/filecfg\_ & 6.3 & 6.3 & 6.3 & 6.3 & 2.4 & 2.4 & 2.4 & 2.4 \\
14 & draco/draco\_me & 8.4 & 8.4 & 4.8 & 18 & 7.3 & 8.1 & 3.4 & 17 & 64 & ndpi/filecfg\_ & 6.3 & 6.3 & 6.3 & 6.3 & 2.7 & 2.7 & 2.7 & 2.7 \\
15 & draco/draco\_pc & 15 & 3.5 & 5.0 & 7.2 & 13 & 2.9 & 4.0 & 6.7 & 65 & ndpi/filecfg\_ & 6.0 & 6.1 & 6.0 & 6.0 & 1.7 & 1.8 & 1.7 & 1.8 \\
16 & draco/draco\_pc & 15 & 5.1 & 3.6 & 2.9 & 15 & 3.8 & 3.1 & 2.0 & 66 & ndpi/filecfg\_ & 5.8 & 5.8 & 5.8 & 5.8 & 1.9 & 1.9 & 1.9 & 2.1 \\
17 & fftw3/fftw3\_ & 21 & 21 & 21 & 21 & 32 & 33 & 32 & 32 & 67 & ndpi/filecfg\_ & 0.4 & 0.4 & 0.4 & 0.4 & 0.9 & 0.9 & 0.9 & 0.9 \\
18 & freerdp/TestFuzz & 14 & 14 & 14 & 13 & 10 & 10 & 10 & 10.0 & 68 & ndpi/filecfg\_ & 7.8 & 8.0 & 8.0 & 8.0 & 3.8 & 3.9 & 3.9 & 3.8 \\
19 & glslang/compile\_ & 21 & 21 & 21 & 22 & 19 & 19 & 19 & 19 & 69 & ndpi/filecfg\_ & 6.7 & 6.7 & 6.7 & 6.7 & 2.3 & 2.3 & 2.3 & 2.3 \\
20 & harfbuzz/hb-shape & 19 & 19 & 19 & 19 & 23 & 23 & 23 & 23 & 70 & ndpi/process\_ & 47 & 44 & 50 & 45 & 41 & 37 & 43 & 38 \\
21 & harfbuzz/hb-subse & 3.1 & 3.2 & 2.9 & 3.0 & 2.0 & 2.1 & 1.7 & 1.9 & 71 & opencv/imread\_ & 4.7 & 4.1 & 4.2 & 4.7 & 3.5 & 2.9 & 3.1 & 3.5 \\
22 & hwloc/hwloc\_ & 14 & 14 & 14 & 14 & 61 & 62 & 63 & 62 & 72 & openexr/openexr\_ & 5.1 & 5.2 & 5.3 & 5.0 & 4.3 & 4.2 & 4.3 & 4.1 \\
23 & icu/calendar & 16 & 16 & 16 & 16 & 12 & 12 & 12 & 12 & 73 & openexr/openexr\_ & 5.2 & 4.8 & 4.9 & 4.8 & 4.5 & 4.0 & 4.1 & 4.1 \\
24 & icu/date\_tim & 14 & 12 & 14 & 12 & 11 & 8.5 & 11 & 8.5 & 74 & openssh/authopt\_ & 0 & 0 & 0 & 3.2 & 0 & 0 & 0 & 3.4 \\
25 & icu/normaliz & 16 & 16 & 16 & 16 & 16 & 17 & 17 & 17 & 75 & openssh/kex\_fuzz & 0 & 0 & 9.5 & 7.2 & 0 & 0 & 10 & 7.8 \\
26 & icu/number\_f & 15 & 15 & 16 & 16 & 12 & 12 & 13 & 13 & 76 & openssh/privkey\_ & 0 & 0 & 6.0 & 4.5 & 0 & 0 & 4.7 & 3.5 \\
27 & icu/unicode\_ & 18 & 13 & 11 & 16 & 15 & 9.8 & 9.5 & 12 & 77 & openssh/sig\_fuzz & 0 & 0 & 0 & 0 & 0 & 0 & 0 & 0 \\
28 & imgmk/encoder\_ & 0.8 & 0.8 & 0.8 & 0.8 & 0.7 & 0.7 & 0.7 & 0.7 & 78 & openssh/sshsigop & 0 & 0 & 0 & 0 & 0 & 0 & 0 & 0 \\
29 & imgmk/ping\_ & 5.5 & 5.4 & 5.0 & 5.0 & 4.3 & 4.2 & 3.9 & 3.9 & 79 & openssl/acert & 2.1 & 2.2 & 2.2 & 2.1 & 1.5 & 1.6 & 1.6 & 1.5 \\
30 & iperf/cjson\_ & 24 & 25 & 25 & 25 & 26 & 28 & 27 & 27 & 80 & openssl/asn1pars & 2.2 & 2.2 & 2.2 & 2.2 & 1.6 & 1.6 & 1.5 & 1.6 \\
31 & jq/jq\_parse & 3.5 & 3.5 & 3.5 & 3.5 & 3.6 & 3.6 & 3.6 & 3.6 & 81 & openssl/cms & 2.4 & 2.4 & 2.4 & 2.4 & 1.8 & 1.8 & 1.8 & 1.8 \\
32 & jq/jq\_parse & 6.5 & 6.5 & 7.3 & 6.4 & 6.4 & 6.5 & 7.2 & 6.3 & 82 & openssl/v3name & 2.2 & 2.8 & 0 & 2.7 & 1.7 & 2.2 & 0 & 2.1 \\
33 & libcoap/get\_asn1 & 0.5 & 0 & 0 & 0 & 0.3 & 0 & 0 & 0 & 83 & php/-json & 0 & 5.2 & 0 & 0 & 0 & 1.4 & 0 & 0 \\
34 & libcoap/oscore\_c & 1.9 & 0 & 0 & 0 & 1.1 & 0 & 0 & 0 & 84 & php/-unseria & 0 & 0 & 0 & 4.9 & 0 & 0 & 0 & 1.2 \\
35 & libcoap/split\_ur & 2.0 & 0.8 & 0 & 0 & 1.6 & 0.8 & 0 & 0 & 85 & php/-unseria & 0 & 0 & 5.0 & 0 & 0 & 0 & 1.2 & 0 \\
36 & libgit2/objects\_ & 2.3 & 2.5 & 2.3 & 2.3 & 1.9 & 2.1 & 1.8 & 1.8 & 86 & pjsip/crypto & 56 & 17 & 57 & 56 & 63 & 6.9 & 65 & 61 \\
37 & libgit2/patch\_pa & 1.5 & 1.5 & 1.5 & 1.5 & 1.0 & 1.0 & 1.0 & 1.0 & 87 & pjsip/dns & 19 & 15 & 15 & 15 & 12 & 8.5 & 8.7 & 8.7 \\
38 & libical/libical\_ & 13 & 12 & 12 & 14 & 11 & 11 & 11 & 12 & 88 & pjsip/sip & 0 & 18 & 12 & 16 & 0 & 9.7 & 7.2 & 9.2 \\
39 & libical/libicalv & 6.0 & 8.5 & 7.9 & 6.0 & 6.1 & 9.2 & 8.3 & 6.3 & 89 & pjsip/srtp & 22 & 22 & 22 & 16.8 & 11 & 11 & 11 & 8.5 \\
40 & libjxl/cjxl\_ & 21 & 21 & 21 & 21 & 15 & 15 & 15 & 15 & 90 & pugixml/parse & 15 & 15 & 15 & 15 & 15 & 16 & 15 & 15 \\
41 & libjxl/color\_en & 81 & 80 & 80 & 80 & 97 & 96 & 96 & 96 & 91 & pugixml/xpath & 41 & 42 & 41 & 39 & 35 & 35 & 35 & 34 \\
42 & libjxl/fields\_ & 25 & 25 & 25 & 25 & 45 & 45 & 45 & 45 & 92 & quickjs/compile & 28 & 28 & 28 & 28 & 24 & 24 & 25 & 24 \\
43 & libjxl/icc\_code & 7.2 & 7.3 & 7.2 & 7.2 & 17 & 18 & 18 & 18 & 93 & quickjs/eval & 27 & 27 & 26 & 26 & 24 & 24 & 23 & 23 \\
44 & libjxl/set\_from & 16 & 19 & 18 & 21 & 16 & 17 & 17 & 20 & 94 & sswan/crls & 11 & 11 & 11 & 11 & 7.0 & 7.0 & 6.9 & 7.0 \\
45 & libpcap/both & 27 & 24 & 22 & 19 & 30 & 27 & 24 & 22 & 95 & sswan/ids & 9.3 & 8.1 & 9.4 & 7.1 & 5.7 & 5.4 & 5.8 & 4.4 \\
46 & libpcap/filter & 25 & 21 & 25 & 25 & 27 & 25 & 28 & 28 & 96 & wabt/wasm2wat & 60 & 61 & 56 & 56 & 68 & 70 & 64 & 64 \\
47 & libplist/bplist\_ & 9.9 & 14 & 9.9 & 10 & 8.3 & 12 & 8.3 & 8.4 & 97 & yajl/json\_ & 79 & 70 & 78 & 79 & 69 & 64 & 69 & 69 \\
48 & libplist/jplist\_ & 12 & 12 & 12 & 12 & 11 & 10 & 11 & 11 & 98 & zlib/compress & 51 & 51 & 51 & 48 & 41 & 41 & 42 & 38 \\
49 & libplist/oplist\_ & 11 & 11 & 10 & 10 & 8.3 & 8.3 & 8.1 & 8.1 & 99 & zlib/zlib\_unc & 53 & 53 & 54 & 54 & 49 & 49 & 50 & 50 \\
50 & libplist/xplist\_ & 13 & 14 & 11 & 13 & 9.8 & 12 & 7.9 & 10 & 100 & zopfli/zopfli\_d & 86 & 86 & 86 & 86 & 70 & 71 & 70 & 70 \\
\midrule
& \textbf{Avg} & \textbf{16.9} & \textbf{16.2} & \textbf{17.5} & \textbf{17.5} & \textbf{16.6} & \textbf{15.9} & \textbf{17.3} & \textbf{17.3} & \multicolumn{10}{c}{} \\
& \textbf{Prod.} & \multicolumn{1}{c}{84} & \multicolumn{1}{c}{83} & \multicolumn{1}{c}{88} & \multicolumn{1}{c}{89} & & & & & \multicolumn{10}{c}{} \\
& \textbf{Cost} & \multicolumn{1}{c}{\$1.04} & \multicolumn{1}{c}{\$1.35} & \multicolumn{1}{c}{\$1.14} & \multicolumn{1}{c}{\$1.64} & & & & & \multicolumn{10}{c}{} \\
\cmidrule{1-10}
\end{tabular}}%

\footnotesize
\noindent AP = w/o adversarial probe gate; P2 = w/o P2 report; St = w/o static analysis; Gm = Gemini~3.1~Flash. Prod. = productive (harness builds and exercises non-zero target-library lines).
\end{table*}